\documentclass[11pt,epsf]{article}
\usepackage{hyperref}
\usepackage{slashed}
\usepackage{graphicx}
\usepackage{amssymb}
\usepackage{amsmath}
\usepackage{epsf}
\usepackage{rotating}
\usepackage[small]{caption}
\usepackage[numbers,square]{natbib}

\def\a{\alpha}

\def\g{\gamma}

\let\G=\Gamma

\def\beqn{\begin{eqnarray}}
\def\eeqn{\end{eqnarray}}
\def\ba{\begin{eqnarray}}
\def\ea{\end{eqnarray}}
\newcommand{\bea}{\begin{eqnarray}}
\newcommand{\eea}{\end{eqnarray}}

\newcommand{\beqa}{\begin{eqnarray}}
\newcommand{\eeqa}{\end{eqnarray}}
\def\beq{\begin{equation}}
\def\eeq{\end{equation}}

\let\G=\Gamma

 \let\n=\nu 

\begin{document}
\begin{center}
\vspace{5.cm}

{\bf \large 
Cosmological Properties of a Gauged Axion  \\}
\vspace{1.5cm}
{ \bf  $^a$Claudio Corian\`{o}, $^b$Marco Guzzi, $^c$George Lazarides and $^a$Antonio Mariano\\}

\vspace{1cm}

{\it  $^a$Dipartimento di Fisica, Universit\`{a} del Salento and  \\
INFN Sezione di Lecce, Via Arnesano 73100 Lecce, Italy\footnote{claudio.coriano@le.infn.it, antonio.mariano@le.infn.it}\\}
\vspace{.3cm}
{\it $^b$Department of Physics, Southern Methodist University, \\Dallas TX 75275, USA\footnote{mguzzi@physics.smu.edu}\\}
\vspace{.3cm}
{ \it $^c$ Physics Division, School of Technology, \\ Aristotle University of Thessaloniki, \\Thessaloniki 54124, Greece\footnote{lazaride@eng.auth.gr}\\}

\begin{abstract}
We analyze the most salient cosmological features of axions in extensions of the Standard Model with a gauged anomalous extra $U(1)$ symmetry. The model is built by imposing the constraint of gauge invariance in the anomalous effective action, which is extended with Wess-Zumino counterterms. These generate axion-like interactions of the axions to the gauge fields and a gauged shift symmetry. The scalar sector is assumed to acquire a non-perturbative potential after inflation, at the electroweak phase transition, which induces a mixing of the St\"uckelberg field of the model with the scalars of the electroweak sector, and at the QCD phase transition. We discuss the possible mechanisms of sequential misalignments which could affect the axions of these models, and generated, in this case, at both transitions. We compute the contribution of these particles to dark matter, quantifying their relic densities as a function of the St\"uckelberg mass. We also show that models with a single anomalous U(1) in general do not account for the dark energy, due to the presence of mixed $U(1)-SU(3)$ anomalies. 

\end{abstract}
\end{center}
\newpage

\section{Introduction} 
Given its important role as a possible solution of the strong CP problem \cite{Peccei:1977ur}
as well as a candidate for the dark matter of the universe, the study of axions \cite{Wilczek:1977pj,Weinberg:1977ma} \cite{Dine:1981rt,Zhitnitsky:1980tq,Kim:1979if,Shifman:1979if} (see \cite{Sikivie:2006ni} for an overview)
has received momentum both at theoretical and experimental level along the years. 
The invisible axion owes its origin to a global $U(1)_{PQ}$  (Peccei-Quinn, PQ) symmetry which is spontaneously broken in the early universe and explicitly broken to a discrete $Z_N$ symmetry by instanton effects at the QCD phase transition \cite{Sikivie:1982qv}. 
The breaking occurs at a temperature $T_{PQ}$ below which the symmetry is nonlinearly realized. 
Strings and domain walls relics, which are typical of axion models and are a problem in ordinary PQ cosmology, can be avoided by introducing inflation to account for their dilution, or by embedding the model into more general constructions based on theories of Grand Unification \cite{Lazarides:1982tw}.

The almost massless nature of the axion and its suppressed coupling to the fields of the Standard Model are consequences of the fact that this field is associated with the phase of a global anomalous symmetry. Both properties are related to the same scale, the axion decay constant $f_a\sim 10^{10}-10^{12}$ GeV. 

The implications of the PQ axion in cosmology, both in supersymmetric and in non supersymmetric models, have been explored to a finer level of detail. For instance, the axion plays an important role in determining the structure of the primordial perturbations 
\cite{Dimopoulos:2003ii,Dimopoulos:2003ss,Lazarides:2005ek}, where it can act as a curvaton. 

The gauging of an anomalous symmetry has some important effects on the properties of this pseudoscalar, first among all the appearance of independent mass and couplings to the gauge fields. This scenario allows a wider region of parameter space where to look for these particles. For this reason, axion-like fields, which are at the center of several investigations, are unlikely to find any significant and fundamental formulation without an underlying anomalous $U(1)$ gauge symmetry, as emphasized in previous works~\cite{Coriano:2006xh,Coriano:2007xg}, \cite{Coriano:2009zh}. 
 
So far only two complete models have been put forward for a consistent analysis of these types of particles, the MLSOM \cite{Coriano':2005js} and the USSM-A \cite{Coriano:2008xa}. 
The first of them is at the basis of the elaborations that we are going to provide in this work. Here we will be focusing on the phenomenological analysis of a scenario which is a direct consequence of the model introduced in \cite{Coriano':2005js}, while more details on the supersymmetric construction will be discussed in a separate work. 
 
Although the natural framework that motivates these constructions is open string theory \cite{Angelantonj:2002ct}, the effective actions describing these types of particles can be consistently defined at lower energy just by the inclusion of the relevant dimension-5 Wess-Zumino (Peccei-Quinn) interactions. These are necessary in order to guarantee the gauge invariance of the effective action and can be interpreted as counterterms. In fact, they balance the anomalous variation of the 1-loop effective action induced by the extra $U(1)$ symmetry, restoring the gauge symmetry.
 
The gauging of an anomalous symmetry is the essential element in the construction of these effective actions and can be justified within intersecting brane models. 
The gauging is a variant of the standard Peccei-Quinn construction and is characterized by a new scale $M$, which is the St\"uckelberg mass. We recall that St\"uckelberg extensions of the Standard Model with a non-anomalous $U(1)$ have been analyzed in several recent works \cite{Feldman:2007wj,Feldman:2006wb,Feldman:2010wy}.

We are going to provide a physical perspective on the possible phenomenological implications of the anomalous case. In particular, we will try to connect the St\"uckelberg fields, which are in the spectrum of these models, to the physical axion which may appear as an extremely 
weakly interacting particle of a certain relic density in our current universe. Our assumption, in the identification of the physical axion, is that the original St\"uckelberg fields will mix at the electroweak phase transition with the Higgs sector. 
As a result, an almost massless state will emerge after the electroweak phase transition. 
 
One of the key mechanisms that we will try to adapt and extend from the PQ case is that of {\em vacuum misalignment}. This phenomenon occurs whenever a quasi Nambu-Goldstone mode - generated by the breaking of a certain symmetry - acquires non-perturbatively a small potential, lifting one flat direction from the vacuum degeneracy. For axions characterized both by an $SU(2)$ and an $SU(3)$ charge the mechanism of vacuum misalignment becomes sequential, as we are going to show.
 
From a more general perspective, we will also try to characterize the possible role of these types of particles as quintessence axions. These appear in models where the axions remain decoupled from the gluonic sector and their mass is purely of electroweak origin (see for instance \cite{Nomura:2000yk}). In this case one tries to exploit the Nambu-Goldstone nature 
of these particles.  The main idea behind this proposal is that a phase transition around the electroweak scale can generate a small curvature in the potential, capable of giving a tiny mass to this particle, smaller than the Hubble parameter at current time ($H_0$). For this to be possible, as we are going to show, one has to search for solutions of the anomaly equations for an anomalous $U(1)$ which has a vanishing mixed anomaly with the $SU(3)$ color group. In this case the only source of mass for these axions would come from the electroweak and  not from the QCD phase transition, and as such could be extremely small. 
   
In the general models that we analyze, the anomaly equations do not allow for such a solution, although this would not exclude the possibility of finding others, in the presence of more complicated gauge structures, for instance in models with several U(1)'s. 
We will not address this specific point any further, leaving it as an option for future studies. Instead, we will concentrate on the general features of an axion-like field coming from a single anomalous $U(1)$ symmetry, characterized by the presence of mixed anomalies both with the $SU(2)$ and $SU(3)$ sectors. The phenomenological details of the model are rather intricate, and have been worked out before. For this reason we have summarized in the next section some of their salient features, which turn out to be necessary in order to proceed with a realistic estimate of the relic densities. This is the specific goal of our work.

\section{General features of models with gauged axions: the St\"uckelberg field}
\label{sec:gener-feat-models}
In this section we briefly review the main features of the class of models that we address, discussing specifically 
the St\"uckelberg field $b$ which accompanies their anomalous $U(1)_B$ symmetry. It has been included in order to clarify the origin of the anomalous gauging and to compare the roles played by the PQ ($a$) and the St\"uckelberg axions, which is relevant for the analysis that will follow. The structure of the entire Lagrangian is discussed in \cite{Coriano':2005js} and has been briefly summarized, in part,  in the appendix.

Intersecting brane models are one of those constructions where these types of generalized axions appear \cite{Ibanez:2001nd,Antoniadis:2002qm,Blumenhagen:2006ci}. In the case in which several stacks of branes are introduced, each stack being the domain in which fields with the gauge symmetry $U(N)$ live, 
several intersecting stacks generate, at their common intersections, fields with the quantum numbers of all the unitary gauge groups of the construction, such as $U(N_1)\times U(N_2)\times ...\times U(N_k)=SU(N_1)\times U(1)\times SU(N_2)\times U(1)\times  ...\times SU(N_k)\times U(1)$. In realistic models, the phases of the extra $U(1)$'s are rearranged in terms of an anomaly-free generator, with an (anomaly free) hypercharge U(1) (or $U(1)_Y$) times extra 
$U(1)$'s which are anomalous, carrying both their own anomalies and the mixed anomalies with all the fields of the Standard Model.  

For instance, a simple realization of the Standard Model is obtained by taking 3 stacks of branes: a first stack of 3 branes, with a symmetry $U(3)$, a second stack of 2 branes, with a symmetry 
$U(2)$ and an extra single brane $U(1)$, giving a gauge structure of the form $SU(3)\times SU(2)\times 
U(1)\times U(1)\times U(1)$. Linear combinations of the generators of the three $U(1)$'s allow to rewrite the entire abelian symmetry in the form $U(1)_Y\times U(1)' \times U(1)''$. These rearrangements of the $U(1)$ phases have been studied in the previous literature. For instance, the original basis for the U(1)'s  is also called ``the brane basis'',  while the reorganization of the generators in the form of ``hypercharge plus reminder'' goes under the name of ``the hypercharge basis''. 
There are explicit assignments in the recent literature  \cite{Ibanez:2001nd,Antoniadis:2002qm,Leontaris:2007ej}.

The simplest realization of the Standard Models (SM) is obtained by 2 stacks and a single brane at their intersections, giving a symmetry $U(3)\times U(2)\times U(1)$. In this case, in the hypercharge basis, the gauge structure of the model can be rewritten in the form $SU(3)_c\times SU(2)_w\times U(1)_Y\times U(1)' \times U(1)''$. We will be using also the notation $U(1)_B\times U(1)_C$ to refer to the two 
$U(1)$ factors ($U(1)'\times U(1)''$) of the abelian gauge structure. As often emphasized in previous works, the 
two extra U(1)'s are in a ``broken'' phase. For instance, if we denote with $B$ and $C$, the kinetic terms of these abelian anomalous gauge fields are given by
\ba
\mathcal{L}_{St}=\frac{1}{2}\left( \partial_\mu b - M_1 B_\mu\right)^2 + \frac{1}{2}\left( \partial_\mu c - M_2 C_\mu\right)^2, 
\label{stuclag}
\ea
which is the well-known St\"uckelberg form. $M_1$ and $M_2$ are also called St\"uckelberg masses while $b$ and 
$c$ are two pseudoscalars known as St\"uckelberg fields (or ``St\"uckelberg axions''). The St\"uckelberg symmetry of the Lagrangian (\ref{stuclag}) is revealed by acting with gauge transformations of the gauge fields $B$ and $C$, under which their corresponding axions $b$ and $c$ vary by a local shift 
\ba
&&\delta_B B_\mu=\partial_\mu \theta_B \qquad \qquad \delta b= M_1 \theta_B \nonumber \\
&&\delta_C C_\mu=\partial_\mu \theta_C \qquad \qquad \delta c= M_2 \theta_C, \nonumber
\ea
parameterized by the local gauge parameters $\theta_B$ and $\theta_C$.
In the literature, the St\"uckelberg symmetry is presented as a way to give a mass to an abelian gauge field but still preserving the gauge symmetry of the theory. However, a more careful look at this symmetry shows that its realization is 
the same one obtained, for instance, in an abelian Higgs model when one decouples the radial excitations of the Higgs fields from its phase \cite{Coriano:2009zh}. Therefore, in this respect, the symmetry does not appear to contain much novelty. However, in the effective theory which characterizes these models, the mechanism which generates the mass of the anomalous $U(1)$'s is unrelated to the traditional Higgs mechanism, since there is no Higgs potential involved. 

The massive anomalous gauge bosons acquire a mass through the presence of 
``$A\wedge F\,$'' couplings in the effective string theory description (see for instance \cite{Ghilencea:2002da}). The starting Lagrangian of the effective theory involves an antisymmetric  rank-2 tensor $A_{\mu\nu}$ coupled to the field strength $F_{\mu\nu}$ of an anomalous gauge boson 
(here denoted by $B$)
\begin{equation}
\label{dd} 
{\cal L}\ =\ -\frac{1}{12} H^{\mu\nu\rho}
H_{\mu\nu\rho}-\frac{1}{4g^2} F^{\mu\nu} F_{\mu\nu} 
+ \frac{M}{4}\ \epsilon^{\mu\nu\rho\sigma} { A}_{\mu\nu}\ F_{\rho\sigma},
\end{equation}
where 
\beq
\label{field3}
H_{\mu\nu\rho}=\partial_\mu A_{\nu\rho}+\partial_\rho A_{\mu\nu}
+\partial_\nu A_{\rho\mu}, \qquad  F_{\mu\nu}=\partial_\mu
B_\nu-\partial_\nu B_\mu
\eeq
is the kinetic term for the 2-form and $g$
is an arbitrary constant. Beside the two kinetic terms for $A_{\mu\nu}$ and $B_\mu$, the third contribution in Eq.~(\ref{dd}) is 
the $A\wedge F$ interaction. 

The Lagrangian is dualized by using a ``first order'' formalism, where $H$ is treated independently from the antisymmetric field $A_{\mu\nu}$. This is obtained by introducing a constraint with a Lagrangian multiplier field $b(x)$ in order to enforce the condition $H=dA$ from the equations of motion of $b$, in the form  
\begin{equation}
\label{secondform}
{\cal L}_0=-\frac{1}{12} H^{\mu\nu\rho} H_{\mu\nu\rho}-\frac{1}{4g^2} F^{\mu\nu}\ F_{\mu\nu} 
- \frac{M}{6}\ \epsilon^{\mu\nu\rho\sigma} H_{\mu\nu\rho}\ B_{\sigma} 
+\frac{1}{6}\,b(x)\,\epsilon^{\mu\nu\rho\sigma} \partial_\mu H_{\nu\rho\sigma}.
\end{equation}
The appearance of a scale $M$ in this Lagrangian is of paramount importance both in the analysis of the relic densities of axions generated by the dualization of this action, and in determining the mass of the extra anomalous U(1) gauge boson, which has been analyzed in detail in previous works \cite{Armillis:2008vp}. It defines the energy region where the Green-Schwarz mechanism comes into play to cancel the anomaly in orientifold vacua of string theory \cite{Coriano':2005js}.  Clearly, it is part of a far more involved field theory Lagrangian which, in general, is not included in the field theory analysis of this mechanism, since the expansion stops at operators of dimension 5. We just remark, at this point, that the appearance of the St\"uckelberg description in theories with gauge anomalies is not limited to effective field theories derived from strings, but it is also common to simple 2-dimensional models, such as the bosonized Schwinger model (see \cite{Aurilia:1980jz}).

The last term in (\ref{secondform}) is necessary in order to reobtain (\ref{dd}) from (\ref{secondform}). If, instead, we integrate by parts the last term of the Lagrangian given in (\ref{secondform}) and solve trivially for $H$ we find 
\begin{equation}
H^{\mu\nu\rho}= - \epsilon^{\mu\nu\rho\sigma}\left(M B_\sigma-\partial_\sigma b\right).
\end{equation}
Inserting  this back into (\ref{secondform}) we obtain the expression
\begin{equation}
{\cal L}_{A}\ =\ -\frac{1}{4g^2}\ F^{\mu\nu}\ F_{\mu\nu} - \frac{1}{2} \left(M B_\sigma-\partial_\sigma b\right)^2
\end{equation}
which is the St\"uckelberg form for the mass terms of $B$.\\
This rearrangement of the degrees of freedom, valid in a classical sense \cite{Duff:1980qv}, and the mapping of the possible physical phases of these 
two model theories, is an example of the connection between Lagrangians of antisymmetric tensor fields and their dual formulations, that in this specific case is an abelian massive Yang-Mills theory in a St\"uckelberg form (see for instance the discussion in \cite{Quevedo:1996uu}).  

The axion field generated by the dualization mechanism appears to be a Nambu-Goldstone mode, which could be absorbed by a  unitary gauge choice in the (defining) St\"uckelberg phase of the model. However, as discussed in \cite{Coriano':2005js}, we will allow a mixing between this mode and the Higgs sector at the electroweak phase transition, by introducing an extra potential which respects the gauge symmetry and whose origin has been left, so far, unspecified. This mixing potential is here assumed to be of non-perturbative origin and triggered at the electroweak phase transition. It is parameterized by constants ($\lambda_i$) which are strongly suppressed by the exponential factor  ($\sim e^{-S_{inst}}$, with $S_{inst}$ the instanton action), determined by the value of the action on the instanton background (for electroweak instantons). We will 
come to discuss these points rather closely in the next sections. 

For this reason, at low energy, the counting of the physical degrees of freedom in the pseudoscalar sector of the model is performed in the combined Higgs-St\"uckelberg phase, where a massive physical axion emerges from the combination of the phases of the Higgses and of the St\"uckelberg field. 
In models with several $U(1)$'s  this construction is slightly more involved, but the result of the mixing of the 
complex CP odd phases leaves as a remnant, also in this case, a physical axion, denoted by $\chi$ \cite{Coriano':2005js}, whose mass is controlled by the size of the Higgs-axion mixing. 

The St\"uckelberg Lagrangian that we have reviewed is part of the classical action $S_0$ which also includes the remaining gauge kinetic terms of the theory at classical level, for a symmetry $SU(3)\times SU(2)\times U(1)_Y$. The remaining interactions can be found in the appendix.

\subsection{Charge assignments and counterterms}
We refer to the appendix for more details concerning this class of models and for our conventions, together with a brief outline of the structure of the counterterms in the effective Lagrangian. Here we briefly comment on the list of the charge assignments of the single extra U(1) model, which is given in Table (\ref{solve_q}). 

\begin{figure}[t]
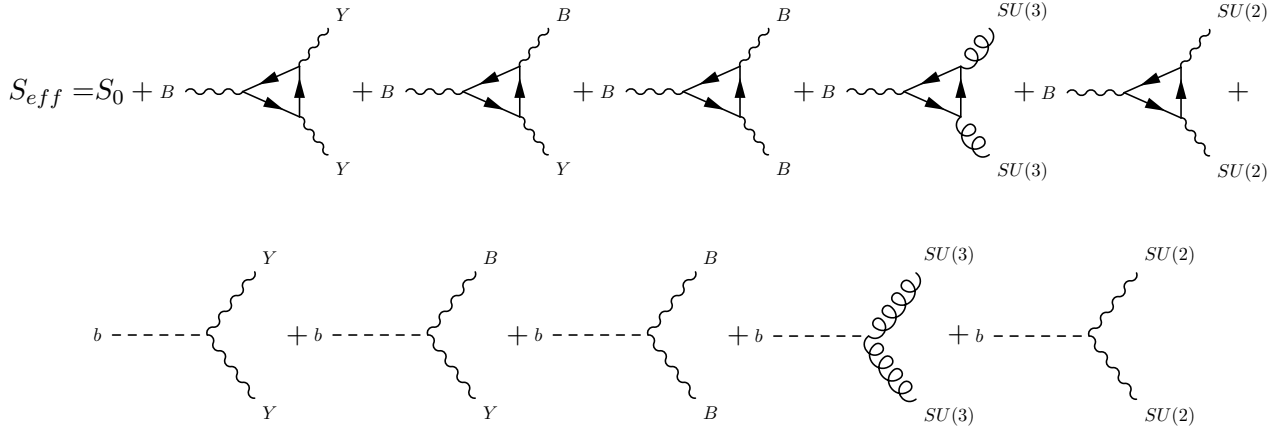

\begin{align}
S_{eff}=&S_0+
\begin{minipage}[c]{70pt}
\includegraphics[scale=.7]{BYY.epsi}
\end{minipage}+
\begin{minipage}[c]{70pt}
\includegraphics[scale=.7]{BYB.epsi}
\end{minipage}+
\begin{minipage}[c]{70pt}
\includegraphics[scale=.7]{BBB.epsi}
\end{minipage}+
\begin{minipage}[c]{70pt}
\includegraphics[scale=.7]{BGG.epsi}
\end{minipage}+
\begin{minipage}[c]{70pt}
\includegraphics[scale=.7]{BWW.epsi}
\end{minipage}+\nonumber\\
&\nonumber\\
&
\begin{minipage}[c]{70pt}
\includegraphics[scale=.7]{axYY.epsi}
\end{minipage}+
\begin{minipage}[c]{70pt}
\includegraphics[scale=.7]{axYB.epsi}
\end{minipage}+
\begin{minipage}[c]{70pt}
\includegraphics[scale=.7]{axBB.epsi}
\end{minipage}+
\begin{minipage}[c]{70pt}
\includegraphics[scale=.7]{axGG.epsi}
\end{minipage}+
\begin{minipage}[c]{70pt}
\includegraphics[scale=.7]{axWW.epsi}
\end{minipage}\nonumber
\end{align}
\caption{Anomalous contributions to the Lagrangian and WZ counterterms\label{fig:lagrangian}}
\end{figure}

Specifically, $ q^B_L, q^B_Q$ denote the charges of the left-handed lepton doublet $(L)$ and of the quark doublet $(Q)$, while $q^B_{u_r},q^B_{d_r}, q^B_{e_R}$ are the charges of the right-handed $SU(2)$ singlets (quarks and leptons). We denote with $\Delta q^B=q^B_u - q^B_d$ the difference between the two charges of the up and down Higgses $(q^B_u, q^B_d)$ respectively.
The trilinear anomalous gauge interactions induced by the anomalous $U(1)$ and the relative counterterms, which are all parts of the 1-loop effective action, are illustrated in Fig. \ref{fig:lagrangian}.
The numerical values of the counterterms appearing on the second line of Fig.~\ref{fig:lagrangian} are fixed by the conditions of gauge invariance of the Lagrangian and are summarized by the following relations
\begin{align}
&C_{BYY} =  -\frac{1}{6}q^B_Q+\frac{4}{3}q^B_{u_R}+\frac{1}{3}q^B_{d_R}-\frac{1}{2}q^B_L+q^B_{e_R},
\nonumber\\
&C_{YBB} = -(q^{B}_Q)^2+2 (q^{B}_{u_r})^2-(q^{B}_{d_R})^2+(q^{B}_L)^2-(q^{B}_{e_R})^2,
\nonumber\\
&C_{BBB} = -6 (q^B_Q)^3+3 (q^B_{u_R})^3+3 (q^B_{d_R})^3-2 (q^B_L)^3+(q^B_{e_R})^3,
\nonumber\\
&C_{Bgg}=\frac{1}{2} (-2 q^B_Q+q^B_{d_R}+q^B_{u_R}),
\nonumber\\
&C_{BWW} = \frac{1}{2} (-q^B_L-3 q^B_Q).
\end{align}
They are, respectively, the counterterms for the cancellation of the mixed anomaly $U(1)_B U(1)_Y^2$ and $U(1)_Y U(1)_B^2$; the 
counterterm for the $BBB$ anomaly vertex or $U(1)_B^3$ anomaly, and those of the $U(1)_BSU(3)^2$ and $U(1)_B SU(2)^2$ anomalies. They are defined in the appendix. From the Yukawa couplings we get the following constraints on the $U(1)_B$ charges
\ba
q^B_Q-q^B_{d}-q^B_{d_R}=0\hspace{1cm}
q^B_Q+q^B_{u}-q^B_{u_R}=0\hspace{1cm}
q^B_L-q^B_{d}-q^B_{e_R}=0.
\ea
In Tab. (\ref{solve_q}) we also show the expressions of the free $U(1)_B$ charges appearing on each generation, having taken into account the conditions of gauge invariance of the Yukawa couplings.  
Using the equations above, we can eliminate some of the charges in the expression of the counterterms, obtaining
\begin{align}
&C_{BYY} = \frac{1}{6} (3 q^B_L+9 q^B_Q+8 \Delta q^B),
\nonumber\\
&C_{YBB} = 2 \left[q^B_d (q^B_L+3 q^B_Q)+2 \Delta q^B (q^B_d+q^B_Q)+(\Delta q^B)^2\right],
\nonumber\\
&C_{BBB} = (q^B_L-q^B_d)^3+3 (q^B_d+q^B_Q+\Delta q^B)^3+3 (q^B_Q-q^B_d)^3-2 (q^B_L)^3-6 (q^B_Q)^3,
\nonumber\\
&C_{Bgg}=\frac{\Delta q^B}{2},
\nonumber\\
&C_{BWW} = \frac{1}{2} (-q^B_L-3 q^B_Q).
\label{charge_asynew}
\end{align}
The solutions given above are generic, in the sense that they parameterize, in principle, an infinite class of models whose charge assignments under 
$U(1)_B$ are arbitrary, with the charges on the last column of Tab. (\ref{solve_q}) taken as their free parameters. One can immediately observe that, due to the presence, in general, of a nonvanishing mixed anomaly of the $U(1)_B$ with both $SU(2)$ and $SU(3)$, the St\"uckelberg axion of the model has interactions both with the strong and the weak sectors, which support instanton solutions, and therefore could acquire a mass non-perturbatively both at the electroweak and at the QCD phase transitions. 
Notice, in particular, that for a model in which $\Delta q=0$, in which both doublets of the Higgs 
sector, $H_u$ and $H_d$ carry the same charge under $U(1)_B$, then the axion mass will not acquire any instanton correction at the QCD phase transition. At the same time, however, it is easy to show that in this case the potential responsible for Higgs-axion mixing disappears. Therefore the axion remains a Nambu-Goldstone mode which is completely absorbed at the electroweak phase transition. In this case, obviously, there is no mechanism of vacuum misalignment for the axion field ($b$). This will contribute to the mass of the two neutral gauge bosons $Z$ and $Z'$, just like all the neutral components of the two Higgses of the model. 

The solution of the same equations with a vanishing electroweak interactions of the St\"uckelberg appears instead possible by choosing $q^B_L= -3 q^B_Q$. In the presence of both a weak ($C_{BWW}$) and a strong ($C_{Bgg}$) counterterm, we will assume that the massless St\"uckelberg 
field $b$ will mix with the scalar CP-odd sector and a physical axion ($\chi$) will emerge from this mixing with a tiny mass ($m_\chi$) generated by electroweak instantons.  The corresponding potential will be rather shallow and for this reason this new degrees of freedom will be essentially 
misaligned but frozen. Its contribution to the relic density will be indeed negligible and for this reason at this stage $\chi$ is extremely light, and massless for all practical purposes. However, due to the presence of a coupling of this field with the strong sector, its mass will be significantly modified at the QCD phase transition, as in the Peccei-Quinn case, with a value which will depend on the size of the St\"uckelberg mass $M$. 

\begin{table}[t]
\begin{center}
\begin{tabular}{|c|c|c|c|c|c|c|}
\hline
$ f $ & $Q$ &  $ u_R $ &  $ d_R $ & $ L $ & $e_R$ \\
\hline \hline
$q^B$  &  $q^B_Q$  & $q^B_{u_R}$  &  $q^B_{d_R}$ & $q^B_L$ & $q^B_{e_R}$ \\ \hline
\end{tabular}
\end{center}
\centering
\renewcommand{\arraystretch}{1.2}
\begin{tabular}{|c|c|c|c|c|}\hline
$f$ & $SU(3)^{}_C$ & $SU(2)^{}_L$ & $U(1)^{}_Y$ & $U(1)^{}_B$ \\ \hline \hline
$Q$ &  3 & 2 & $ 1/6$ & $q_Q^B$\\
$u_R$ &  3 & 1 & $ 2/3$ & $q_Q^B+q^B_{u}$\\
$d_R$ &  3 & 1 & $ -1/3$ & $q_Q^B-q^B_{d}$\\
$L$ &  1 & 2 & $ -1/2$ & $q^B_L $\\ 
$e_R$ &  1 & 1 & $-1$ & $ q^B_L - q^B_{d}$\\ 
\hline
$H^{}_u$ &  1 & 2 & $1/2$ & $ q^B_u $\\ 
$H^{}_d$ &  1 & 2 & $1/2$ & $ q^B_d  $\\ 
\hline
\end{tabular}
\caption{\small Charges of the fermion and of the scalar fields \label{solve_q}}
\end{table}

\section{The electroweak potential for massless fields} 
As in previous works \cite{Coriano:2007xg},  in the construction of the effective action  we follow a bottom-up approach with general charge assignments parameterized just by the set of free charges of $U(1)_B$. These are shown in Fig. \ref{solve_q}, together with the fundamental gauge structure of the Standard Model. The scalar sector of the anomalous abelian models that we are interested in is characterized by a rather standard electroweak potential  involving, in the simplest formulation, two Higgs doublets $V_{P Q}(H_u, H_d)$ plus one extra contribution, denoted as $V_{\slashed{P}\slashed{Q}}(H_u,H_d,b)$ or $V^\prime$, \cite{Coriano':2005js}  which mixes the Higgs sector with the St\"uckelberg axion $b$, needed for the restoration of the gauge invariance of the effective Lagrangian
\begin{equation}
V=V_{PQ}(H_u,H_d) + V_{\slashed{P}\slashed{Q}}(H_u,H_d,b).
\end{equation}
The appearance of the physical axion in the spectrum of the model takes place after 
that the phase-dependent terms, here assumed to be of non-perturbative origin and generated at the electroweak phase transition, find their way in the dynamics of the model and induce a curvature on the scalar potential. The mixing induced in the CP-odd sector determines the presence of a linear combination of the St\"uckelberg field $b$ and of the Goldstones of the CP-odd sector, called $\chi$, which is characterized by an almost flat direction. To better illustrate this point, we begin our analysis by turning to the ordinary potential of 2 Higgs doublets,
\ba
V_{PQ}&=&\mu_u^2 H_u^{\dagger}H_u+\mu_d^2 H_d^{\dagger}H_d+\lambda_{uu}(H_u^{\dagger}H_u)^2
+\lambda_{dd}(H_d^{\dagger}H_d)^2-2 \lambda_{ud}(H_u^{\dagger}H_u)(H_d^{\dagger}H_d)
+2\lambda^{\prime}_{ud}\vert H_u^T \tau_2 H_d\vert^2 \nonumber \\
\eeqa
to which we add a second term
\ba
V_{\slashed{P}\slashed{Q}}&=&\lambda_0(H_u^{\dagger}H_d e^{-i g_B (q_u-q_d)\frac{b}{2 M}})+
\lambda_1(H_u^{\dagger}H_d e^{-i g_B (q_u-q_d)\frac{b}{2 M}})^2+\lambda_2(H_u^{\dagger}H_u)(H_u^{\dagger}H_d e^{-i g_B (q_u-q_d)\frac{b}{2M}})+\nonumber\\
&&\lambda_3(H_d^{\dagger}H_d)(H_u^{\dagger}H_d e^{-i g_B (q_u-q_d)\frac{b}{2 M}})+\textrm{h.c.},
\ea
These terms are allowed by the symmetry of the 
model and are parameterized by one dimensionful  ($\lambda_0$) and three dimensionless constants ($\lambda_1,\lambda_2,\lambda_3$). They are assumed to be generated at the electroweak phase transition non-perturbatively, and as such their values are related to an exponential factor containing as a suppression the instanton action. In the equations below we will rescale $\lambda_0$ by the electroweak scale $v=\sqrt{v_u^2 + v_d^2}$  ($\lambda_0 \equiv \bar{\lambda}_0 v$) so to obtain a homogeneous expression of the mass of $\chi$ as a function of the relevant scales of the model which are, beside the electroweak vev $v$, the St\"uckelberg mass $M$ and the anomalous gauge coupling of the $U(1)_B$, $g_B$.

The physical axion $\chi$ emerges as a linear combination of the phases of the various terms, which are either due to the components of the Higgs sector or to the St\"uckelberg field $b$. To illustrate the appearance of a physical direction in the phase of the extra potential, we focus our attention just on the CP-odd sector of the total potential, which is the only one that is relevant for our discussion.  The expansion of this potential around the electroweak vacuum is given by the parameterization 
\ba
H_u=\left(
\begin{tabular}{c}
$H_u^+$\\
$v_u+H_u^0$
\end{tabular}
\right)
\hspace{1cm}
H_d=\left(
\begin{tabular}{c}
$H_d^+$\\
$v_d+H_d^0$
\end{tabular}
\right).
\ea
This potential is characterized by two null eigenvalues corresponding to two neutral Goldstone modes $(G_0^1,G_0^2)$
and an eigenvalue corresponding to a massive state with an axion component ($\chi$). In the 
$(\textrm{Im}H_d^0,\textrm{Im} H_u^0, b)$ CP-odd basis we get the following normalized eigenstates
\begin{align}
G_0^1&=\frac{1}{\sqrt{v_u^2+v_d^2}}(v_d,v_u,0)\nonumber\\
G_0^2&=\frac{1}{\sqrt{g_B^2 (q_d-q_u)^2 v_d^2 v_u^2+2 M^2 \left(v_d^2+v_u^2\right)}}\left(-\frac{ g_B (q_d-q_u)v_d v_u^2}{\sqrt{v_u^2+v_d^2}},\frac{g_B (q_d-q_u)v_d^2 v_u}{\sqrt{v_d^2+v_u^2}},\sqrt{2}M\sqrt{v_u^2+v_d^2}\right)\nonumber\\
\chi&=\frac{1}{\sqrt{g_B^2 (q_d-q_u)^2 v_u^2 v_d^2+2M^2(v_d^2 + v_u^2)}}
\left(\sqrt{2} M v_u,-\sqrt{2} M v_d, g_B (q_d-q_u) v_d v_u\right)
\end{align}
and we indicate with $O^{\chi}$ the orthogonal matrix which allows to rotate them on the physical basis
\ba
\begin{pmatrix}
G_0^1 \cr
G_0^2 \cr
\chi
\end{pmatrix}
= O^\chi
\begin{pmatrix}
\textrm{Im}H^0_d \cr
\textrm{Im}H^0_u \cr
b
\end{pmatrix},
\ea
which is given by 
\begin{equation}
O^\chi=
\begin{pmatrix}
\frac{v_d}{v} & \frac{v_u}{v} & 0 \cr
-\frac{g_B (q_d-q_u)v_d v_u^2}{v\sqrt{g_B^2 (q_d-q_u)^2 v_d^2 v_u^2+2 M^2 v^2}}&
\frac{g_B (q_d-q_u)v_d^2 v_u}{v\sqrt{g_B^2 (q_d-q_u)^2 v_d^2 v_u^2+2 M^2 v^2}} &
\frac{\sqrt{2}M v}{\sqrt{g_B^2 (q_d-q_u)^2 v_d^2 v_u^2+2 M^2 v^2}} \cr
\frac{\sqrt{2} M v_u}{\sqrt{g_B^2 (q_d-q_u)^2 v_u^2 v_d^2+2M^2 v^2}}&
-\frac{\sqrt{2} M v_d}{\sqrt{g_B^2 (q_d-q_u)^2 v_u^2 v_d^2+2M^2 v^2}} &
\frac{g_B (q_d-q_u) v_d v_u}{\sqrt{g_B^2 (q_d-q_u)^2 v_u^2 v_d^2+2M^2 v^2}} 
\end{pmatrix}
\label{stella}
\end{equation}
where $v=\sqrt{v_u^2+v_d^2}$.\\
$\chi$ inherits WZ interaction since $b$ can be related to the physical axion $\chi$ and to the Goldstone modes via this matrix 
\ba
b &=&  O_{13}^{\chi} G_0^1 + O_{23}^{\chi} G_0^2 + O_{33}^{\chi} \chi ,       
\label{rot12}
\ea
or, conversely,
\ba
\chi &=& O_{31}^{\chi} \textrm{Im}H_d + O_{32}^{\chi} \textrm{Im}H_u + O_{33}^{\chi} b.        
\ea
Notice that the rotation of $b$ into the physical axion $\chi$ involves a factor $O_{33}^{\chi}$ which is of order $v/M$.
This carries as a consequence that $\chi$ inherits from $b$ an interaction with the gauge fields which is suppressed by a scale $M^2/v$. This scale is the product of two contributions: a $1/M$ suppression coming from the original Wess-Zumino counterterm of the Lagrangian ($b/M F\tilde{F}$) and a factor $v/M$ obtained by the projection of $b$ into $\chi$ due to $O_\chi$.

More details on the structure of the various operators appearing in this model have been included in an appendix in order to make our treatment self-contained. We have included also a brief discussion of the construction of $g_{\chi\g\g}$,
which is the factor in front of one of the most important counterterms needed in our numerical analysis and which 
controls the decay of the axion into photons. We briefly comment on its structure. 

The final coupling appears as a coefficient in the interaction of the physical axion with two photons 
\beq
g_{\chi\gamma\gamma}\chi F_\gamma\tilde{F_\gamma}
\eeq
and is given by 
\beqa
g^{\chi}_{\g \g}\, = \, \left(  F O^A_{ W_3 \gamma} O^A_{ W_3 \gamma} 
+ C_{YY}  O^{A}_{Y \gamma  } O^A_{Y \gamma  } \right)\,  O^{\chi}_{33}.
\label{gchichi}
\eeqa
It is defined by a combination of matrix elements of the rotation matrices $O^A$ and $O^\chi$, together with some 
counterterms $F$ and $C_{YY}$. $O^A$  is the matrix that rotates the neutral gauge bosons from the interaction to the mass eigenstates after electroweak symmetry breaking and has elements which are $O(1)$, being expressed in terms of ratios of coupling constants. They correspond to mixing angles. The coefficients $F$ and $C_{YY}$ are the $WZ$ counterterms for cancelling the anomalies emerging 
from the $SU(2) U(1)_B^2$ and $U(1)_B U(1)_{Y}^2$ sectors and can be found in the appendix. They are both  suppressed by $1/M$, while the matrix element 
$O^\chi_{33}$, as we have mentioned, is of order $v/M$. Defining $g^2=g_2^2 + g_Y^2$, the expression of this coefficient can be given in the form 
\beq
g^{\chi}_{\g \g}\,= \frac{g_B g_Y^2 g_2^2}{32 \pi^2 M g^2} O^{\chi}_{3\,3}\sum_f\left(- q^B_{f\,L} +q^B_{f \,R} \left(q^Y_{f \,R}\right)^2 - q^B_{f \,L} \left(q^Y_{f \,L}\right)^2\right).
 \eeq 
 
Notice that this expression is cubic in the gauge coupling constants, since factors such as $g_2/g$ and $g_Y/g$ are mixing angles while the factor $1/\pi^2$ originates from the anomaly. Therefore  one obtains a general behaviour for $g^{\chi}_{\g \g}$ of $O(g^3 v/M^2)$, with charges which are, in general, of order unity.

\subsection{Periodicity of the $V^\prime$ potential}
The phase-dependent potential has a well-defined periodicity. To identify the corresponding phase in the Higgs-neutral 
CP-odd sector we introduce a polar parametrization of the neutral components in the broken electroweak phase 
\ba
H_u^0=\frac{1}{\sqrt{2}}\left(\sqrt{2}v_u + \rho_u^0(x) \right) e^{i\frac{F_u^0(x)}{\sqrt{2}v_u}}
\hspace{1cm}
H_d^0=\frac{1}{\sqrt{2}}\left(\sqrt{2}v_d + \rho_d^0(x) \right) e^{i\frac{F_d^0(x)}{\sqrt{2}v_d}},
\ea
where we have introduced the two phases $F_u$ and $F_d$ of the two neutral Higgs fields. 
The potential is periodic with respect to the linear combination of fields
\ba
\theta(x)\equiv\frac{g_B (q_d-q_u)}{2 M}b(x)-\frac{1}{\sqrt{2}v_u} F_u^0(x) +\frac{1}{\sqrt{2}v_d} F_d^0(x),
\ea
and using the matrix $O^{\chi}$ to rotate on the physical basis, the phase describing the periodicity of the potential turns out to be proportional to the physical axion, modulo a dimensionful constant ($\sigma_\chi$)
\ba
\theta(x)\equiv \frac{\chi(x)}{\sigma_\chi},
\ea
where we have defined
\beq
\sigma_\chi\equiv\frac{2  v_u v_d M}{\sqrt{g_B^2 (q_d-q_u)^2 v_d^2 v_u^2 +2 M^2 (v_d^2+v_u^2)}}.
\eeq
Notice that $\sigma_\chi$, in our case, takes the role of $f_a$ of the PQ case, where the angle of 
misalignment is identified by the ratio $a/f_a$, with $a$ the PQ axion. In our case $\sigma_\chi$, however, is of the order of the electroweak scale. This, as we are going to show, has drastic implications on the relic densities of axions generated at this transition.\\
Notice that $\chi$ (or, equivalently, $\theta$) is gauge invariant as one can check quite directly. In fact a $U(1)_B$ infinitesimal gauge transformation with gauge parameter $\alpha_B(x)$ gives 
\beqa
\delta H_u &=& - \frac{i}{2} q_u g_B \alpha_B H_u \nonumber \\ 
\delta H_d &=& - \frac{i}{2} q_d g_B \alpha_B H_d \nonumber \\ 
\delta F_0^u &=& - \frac{v_u}{\sqrt{2}} q_u g_B \alpha_B \nonumber \\
\delta F_0^d &=& -\frac{v_d}{\sqrt{2}}  q_d g_B \alpha_B \nonumber \\
\delta b &=& -M \alpha_B
\eeqa
giving $\delta\theta=0$. The gauge invariance under $U(1)_Y$ can be easily proven by using the invariance of the St\"uckelberg field $b$ and the fact that the hypercharges of the two Higgses are equal. Finally, the invariance under $SU(2)$ is obvious since the linear combination of the phases that define $\theta(x)$ are not touched by the transformation. From the Peccei-Quinn breaking potential we can extract the following periodic potential
\begin{align}
V^\prime=& 4 v_u v_d
\left(\lambda_2 v_d^2+\lambda_3 v_u^2+\lambda_0\right) \cos\left(\frac{\chi}{\sigma_\chi}\right) + 2 \lambda_1 v_u^2 v_d^2 \cos\left(2\frac{\chi}{\sigma_\chi}\right),
\label{extrap}
\end{align}
with a mass for the physical axion $\chi$  given by
\ba
m_{\chi}^2=\frac{2 v_u v_d}{\sigma^2_\chi}\left(\bar{\lambda}_0 v^2 +\lambda_2 v_d^2 +\lambda_3 v_u^2+4 \lambda_1 v_u v_d\right) 
\approx \lambda v^2.
\label{axionmass}
\ea
Notice that, according to our assumption about the origin of the extra potential, this is driven by the combined product of non-perturbative effects, due to the exponentially small parameters  
$(\bar{\lambda_0}, \lambda_1,\lambda_2,\lambda_3)$, with the electroweak vevs of the two Higgses. Notice also the irrelevance of the St\"uckelberg scale $M$ in determining the value of $\sigma_\chi\sim O(v)$ and of $m_\chi$  near the transition region, due to the large suppression factor $\lambda$ in Eq.~(\ref{axionmass}). 
One point that needs to be stressed is the fact that at the electroweak epoch the angle of misalignment generated by the extra potential is parameterized by $\chi/\sigma_\chi$ while the interaction of the physical axion with the gauge fields is suppressed by $M^2/v$. This feature is obviously 
unusual, since in the PQ case both scales are a single scale, the axion decay constant $f_a$.

We will consider in the next sections two possible scenarios, 
the first is the low gravity scenario, where $M^2/v$ is in the TeV region or above, but essentially disconnected from the typical scale appearing in typical PQ axion models ($f_a\sim 10^{12}$ GeV). In the second scenario we will allow a very large value for 
$M^2/v$, of the same order of $f_a$. In this second case we will re-obtain the PQ axion model, with relic densities for $\chi$ which are comparable with those typical of a PQ axion. The appearance of a misalignment 
angle of the form $\chi/\sigma_\chi$, respect to the PQ case ($a/f_a$), is going to have drastic consequences on the relic densities of this particle generated at the electroweak scale, densities which will be found to be negligible. 

At the QCD phase transition a new - much more sizeable - misalignment occurs and the axion mass gets enhanced by the QCD instantons respect to $m_\chi$ given in (\ref{axionmass}), which in this case is typically of $O(\Lambda_{QCD}^2 v/M^2)$. Notice that at the QCD phase transition $M^2/v$ takes the same role of $f_a$ in PQ (in the PQ case $m_a\sim \Lambda_{QCD}^2/f_a)$). 

There are also some crucial points of difference between a gauged axion and the PQ case that require some comment, since they are not so obvious. Notice, in fact, that in the PQ case, if $v_{PQ}$ is larger than the scale of inflation, then the value of the $\theta$ field can be considered essentially homogeneous. We have already mentioned that in the PQ case $\theta$ is a physical field at every physical scale, since it is a Nambu-Goldstone mode of a global $U(1)$ symmetry and as such cannot  be gauged away. In that case the role of the mechanism of vacuum misalignment at the QCD transition is just to provide a mass for this Goldstone mode. 

In our case, instead, $b$ has no potential and is charged under an anomalous gauge symmetry. As such it appears as a longitudinal component of the anomalous gauge boson $B$, above the electroweak scale. This also implies that there is no effect on $b$ due to inflation, being the St\"uckelberg not a physical field at the scale of inflation. Thus, the reappearance  in the CP-odd sector of a component of $b$ as a physical axion, $\chi$, at the electroweak phase transition, implies that this physical component is not a homogeneous field at the electroweak time. Similar types of inhomogeneities are found also in the PQ case, in models characterized by late inflation. In fact, in that case the homogeneity of the axion field beyond the QCD horizon is not guarantee either. 

For this reason, it is 
conceivable that $\chi$ is homogeneous within the electroweak horizon for the same argument, and one can neglect fluctuations of $\chi$ that enter the horizon at later times. Anyhow, even if these fluctuations were included, they are likely to play a minor role respect to other, more significant effects, such as those determined by the size of the St\"uckelberg mass $M$ ($M^2/v$), which has a dominant impact on the value of the relic densities for these types of axions. For this reason we will be leaving aside possible
further corrections due to a non-homogeneity of the $b$ field beyond the electroweak horizon, knowing that variants of this approach could be worked out following the discussion given in \cite{Chang:1998ys}.

\section{Decays of axion-like particles}
The physical axion acquires a non-vanishing coupling with the massive fermions
that is proportional to the rotation matrix $O^{\chi}$ and to the mass of the fermion.
This coupling increases the number of its decay modes and, in particular, induces new channels in its decay rate into gauge bosons, mediated both by fermion loops and by direct Wess-Zumino interactions.  In this section we perform a complete 
study of the decay rate under the assumption that the mass of $\chi$ is in the meV region and below. In particular, in the case of a very light axi-Higgs, the decays into massless vector bosons
are all dominated by the Wess-Zumino contributions, which are far larger than those coming from the fermion loops.  These results will be used 
in the study of the relic densities of this particle which will be presented in the next sections. Here we compare the results of the decay rates for the new axions with those of the  
the PQ axion that we are going to compute from scratch. 

The interaction of the PQ axion with 
photons is given by 
\ba
{\mathcal L}_{int}= \frac{e^2}{32\pi^2} \frac{c^a_{\g\g}}{f_a} a F\tilde{F} + \dots,
\ea
where we denote with $a$ the axion, which is bounded (from astrophysical and cosmological constraints) 
to be between $10^{8}$ GeV $\leq f_a \leq 10^{12}$ GeV. The dots in the previous formula indicate 
terms that are irrelevant for the current analysis. From a general point of view, 
the coefficient $c^a_{\g\g}$ depends upon the Peccei-Quinn charge assignment and also
on the quark-mass ratios induced by its fermion interactions.
The coupling to the fermions is given by 
\ba
{\mathcal L}_{f}= i g_f \frac{m_f}{v_{PQ}} \bar{\psi_f}\g^5 \psi_f,
\ea
where $m_f$ is the mass of the fermion, whose flavor is denoted by $f$, and the coupling
$g_f=Q_{f_R}-Q_{f_L}$ is given in terms of the chiral PQ charges 
($Q^{PQ}_{f\,L,R}$) of each fermion $(f)$. We denote with $v_{PQ}$ the PQ breaking scale,  which can be taken approximately around $10^{15}$ GeV. 
We recall that in the PQ case the corresponding Wess-Zumino interaction is given by
\ba
{\mathcal L}_{a\g\g}= \frac{G_{a\g\g}}{4}a~F^{\mu\nu} \tilde{F}_{\mu\nu}=-G_{a\g\g} a~\vec{E}\cdot \vec{B},
\ea
where $\vec{E}$ and $\vec{B}$ are the electric and magnetic fields respectively, and the coupling $G_{a\g\g}$
is the sum of a model dependent term and of a second term which depends on the ratio of the quark masses 
\ba
G_{a\g\g} = \frac{\alpha_{em}}{2\pi f_a}\left( \sum_f Q^{PQ}_f (Q_f^{em})^2 - \frac{2}{3}\frac{4+z}{1+z}\right),
\ea
where the quark-mass ratio is $z=m_u/m_d$, while the $Q_f^{em}$'s are the e.m. couplings of the photons to the quarks.
Since the coefficient $G_{a\g\g}$ is model dependent, we can have several possibilities. We compute below
the decay rate into two photons in one specific case in which we assume 
\ba
c^{a}_{\gamma\gamma}=\sum_f Q^{PQ}_f (Q_f^{em})^2=0 
\ea
and $z=0.56$.
This choice gives as a decay rate into two photons
\ba
\Gamma_{a\g\g}=\frac{G_{a\g\g}^2}{64\pi}m_a^3 = 1.1 \times 10^{-24} s^{-1} \left(\frac{m_a}{\textrm{eV}}\right)^5,
\ea
\begin{figure}[t]
\begin{center}
\includegraphics[scale=1]{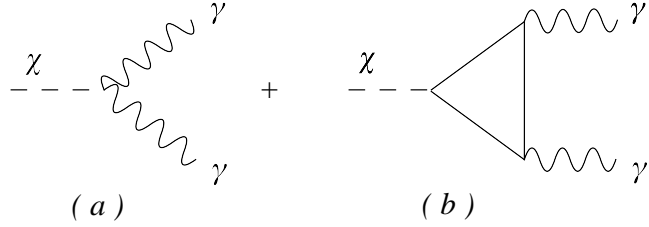}
\caption{ Contributions to the $\chi \rightarrow \g \g$ decay.\label{fig:chi_decay}}
\end{center}
\end{figure}
which is a function of the axion mass $m_a$.
More generally, we want to write the decay rate separating the contribution from the Wess-Zumino interactions  from those which are obtained from the loop corrections. We obtain 
\ba
&&\Gamma_{PQ}(a\rightarrow {\g\g})=\frac{\sum_{spin}|{\mathcal M}_{PQ}|^2}{2m_{a}}\frac{d \vec{k}_1}{(2\pi)^3k_1^0}
\frac{d \vec{k}_2}{(2\pi)^3k_2^0}(2\pi)^4\delta^{(4)}(k-k_1-k_2),
\ea
where the squared amplitude is given by
\ba
&&\sum_{spin}|{\mathcal M}_{PQ}|^2=\sum_{spin}|{\mathcal M}_{point-like} + {\mathcal M}_{loop}|^2
\nonumber\\
&&=8 \left(\frac{c^{a}_{\gamma\gamma}}{F^a}\right)^2 \left(\frac{e^2}{32\pi^2}\right)^2 m_{a}^4 +
\frac{1}{2} \left| \sum_f N_c(f) i \frac{\tau_f ~f(\tau_f)}{4\pi^2 m_f} 
e^2 Q_f^2 \left(g_f \frac{m_f}{v_{PQ}}\right) \right|^2
+\textrm{interf.},
\nonumber\\
\ea

where, in the second term, $N_c(f)$ is the color factor, and the function $\tau_f ~f(\tau_f)$ is a function
of the mass of the fermions circulating in the loop. We have introduced the function $f(\tau)$, defined in any kinematic domain, whose real part is given by 
\ba
\textrm{Re}[f(\tau)] = \left\{ \begin{array}{ll}
(\arcsin{1/\sqrt{\tau}})^2 & \textrm{if}\, \tau \geq 1\\
-\frac{1}{4}\left[\log^2\left(\frac{1+\sqrt{1-\tau}}{1-\sqrt{1-\tau}} \right) -\pi^2 \right] 
& \textrm{if}\, \tau < 1
\end{array} \right.
\ea
while its imaginary part is 
\ba
\textrm{Im}[f(\tau)] = \left\{ \begin{array}{ll}
0 & \textrm{if}\, \tau \geq 1\\
\frac{\pi}{2}\left[\log\left(\frac{1+\sqrt{1-\tau}}{1-\sqrt{1-\tau}} \right)\right] 
& \textrm{if}\, \tau < 1
\end{array} \right.
\ea
where $\tau=4m_f^2/m_{\chi}^2$. In our case we take the branch $\tau > 1$. 

As we move to compute the decay of $\chi$ and assume a free varying mass for this particle, the WZ interaction (Fig.~\ref{fig:chi_decay}a)
\begin{figure}[t]
\begin{center}
\includegraphics[scale=0.3, angle=-90]{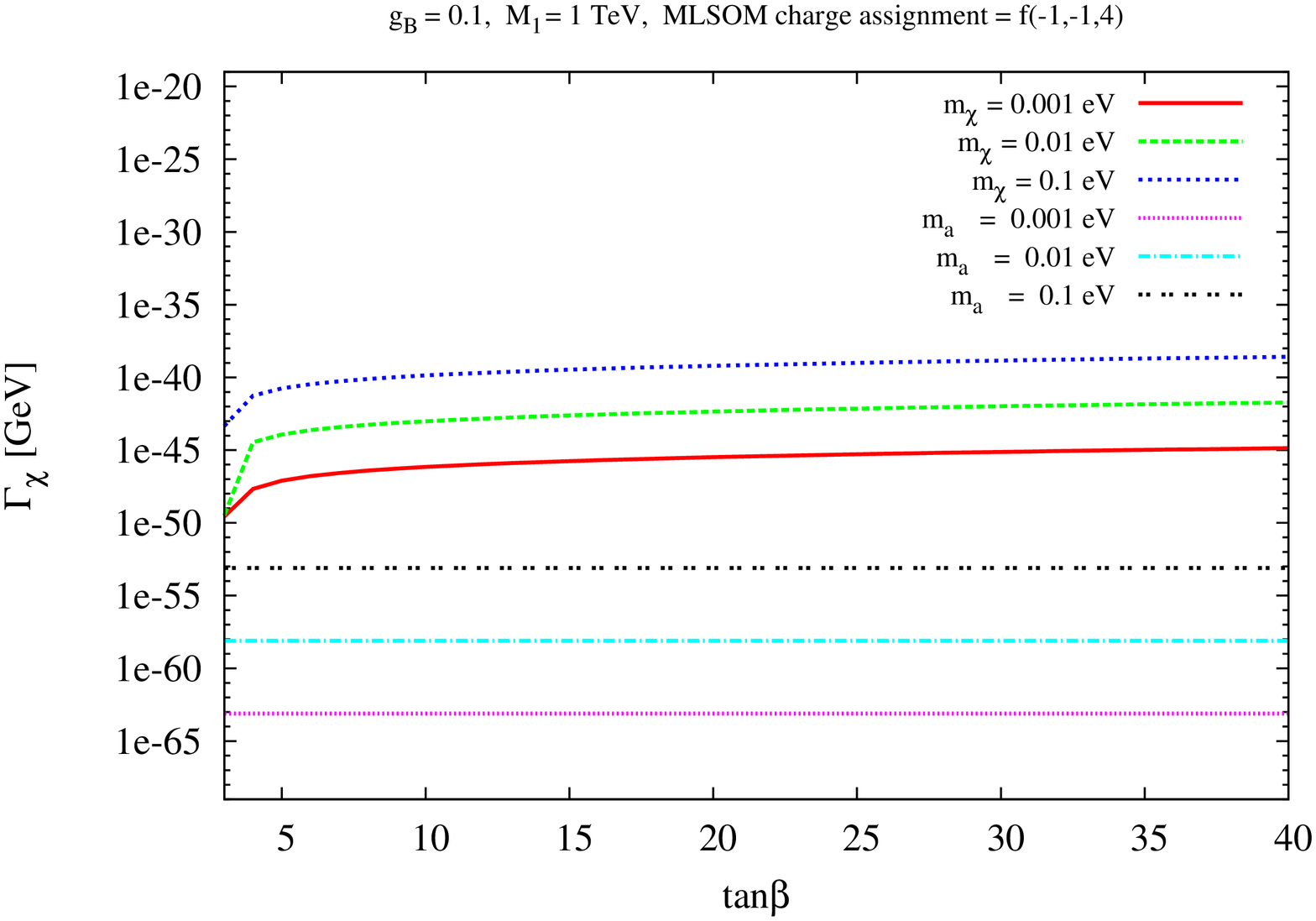}
\includegraphics[scale=0.3, angle=-90]{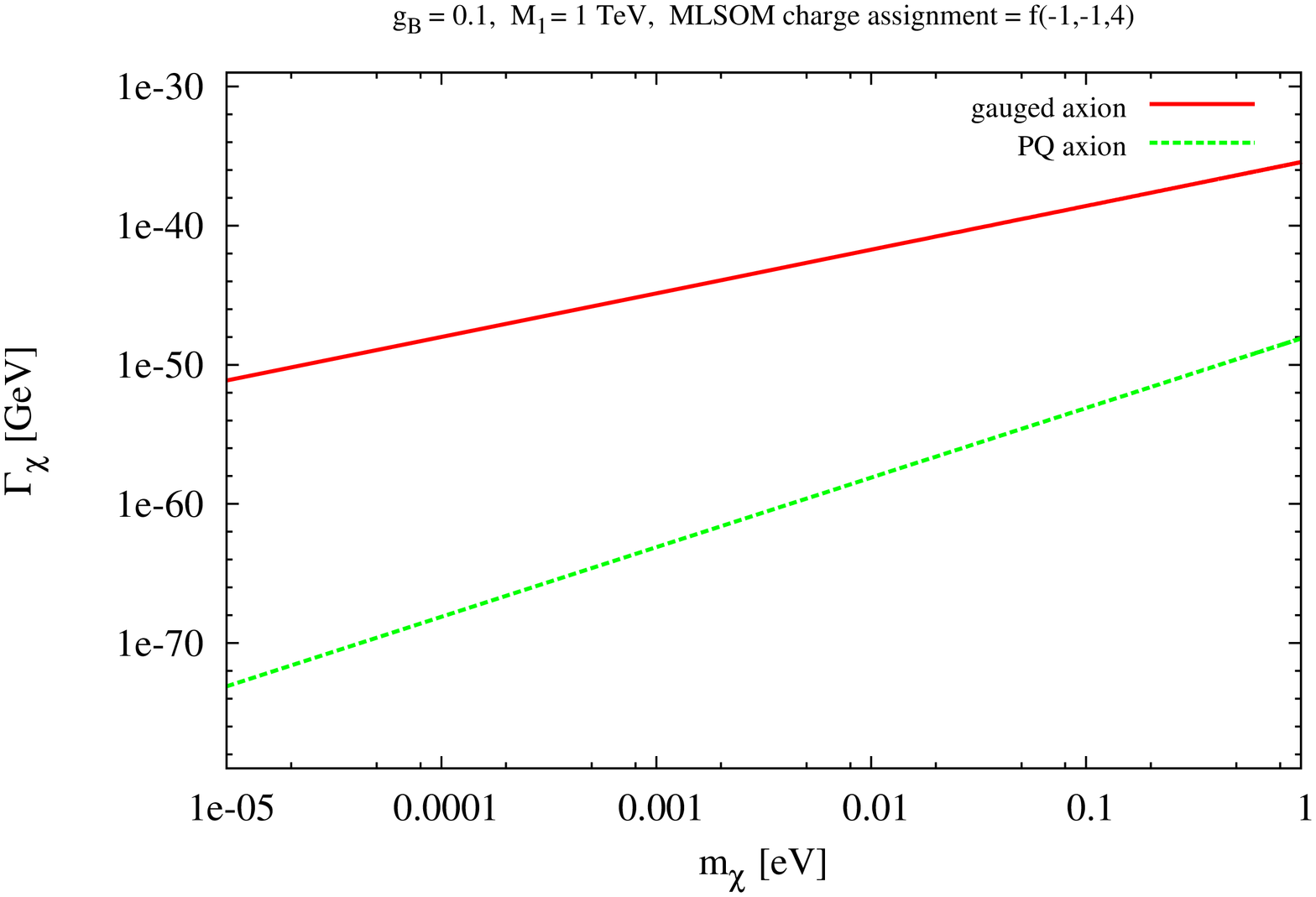}
\end{center}
\caption{\small Total decay rate of the axi-Higgs for several mass values.
Here, for the PQ axion, we have chosen $f_a=10^{10}$ GeV.\label{fig:gamma}}
\end{figure}
is given by
\bea
{\mathcal M}^{\mu \nu}_{WZ}(\chi \rightarrow \g \g) = 4 g^{\chi}_{\g\g} \varepsilon[\mu,\nu,k_1,k_2].
\eea
In Fig.~\ref{fig:chi_decay}a we have isolated the massless contribution to 
the decay rate coming from the WZ counterterm $\chi F^{}_\g \tilde{F}^{}_\g$ whose expression is 
\ba
\label{WZrate1}
\Gamma^{}_{WZ}(\chi \rightarrow \g\g)= \frac{m^3_\chi}{4 \pi}(g^\chi_{\g\g})^2.
\ea

Combining also in this case the tree level decay with the 1-loop amplitude, we obtain for $\chi \rightarrow \g\g$ the amplitude
\ba
{\mathcal M}^{\mu \nu}(\chi \rightarrow \g \g) = {\mathcal M}^{\mu \nu}_{WZ}+{\mathcal M}^{\mu \nu}_{f},
\ea
shown in Fig.~\ref{fig:chi_decay}. In this case the rates are derived from the expression

\ba
\Gamma_\chi\equiv\Gamma(\chi \rightarrow \g\g) &=&  \frac{m^3_\chi}{32 \pi}  \left\{ 8 (g^\chi_{\g\g})^2 
+ \frac{1}{2} \left| \sum_f N_c(f) i \frac{\tau_f ~f(\tau_f)}{4\pi^2 m_f} e^2 Q_f^2 c^{\chi, f} \right|^2  \right.  
\nonumber\\
&& \left. \,\,\,\,\,\,\, \,\,\,\,\,\,\, \,\,\,\,+ \,\,4 g^\chi_{\g\g} 
\sum_f N_c(f) i \frac{\tau_f ~f(\tau_f)}{4\pi^2 m_f} e^2 Q_f^2 c^{\chi, f} \right\}
\ea
and are shown in Fig.\ref{fig:gamma}. In the equation above both the direct ($\sim {(g^\chi_{\g\g}})^2$) and the interference ($\sim g^\chi_{\g\g}$) contributions are suppressed as inverse powers of the St\"uckelberg mass. 
We show the results of this comparative study in Fig. \ref{fig:gamma}, where in the left panel we present results for the decay rates of $\chi\to \g\g$ for several values of the axion mass as a function of $\tan\beta=v_u/v_d$. The plots 
indicate a very mild dependence of the rates on this parameter, even for rather large variations. In the same plot the rates for the PQ case are shown as constant lines, just for comparison. Notice that we have chosen a rather low St\"uckelberg mass, with $M=1$ TeV. The charge assignment of the anomalous model have been denoted as $f(-1,1,4)$, where we have used the convention
\bea
f(q_{Q_L}^B, q_{L}^B, \Delta q^B)\equiv
(q_{Q_L}^B, q_{u_R}^B; q^{B}_{d_R}, q^{B}_{L}, q^{B}_{e_R}, q^{B}_{u},q^{B}_{d}).
\eea
These depend only upon the three free parameters $q^B_{Q_L}$, $q^B_{L},\Delta q^B$. The parametric solution of the anomaly equations of the model $f(q^B_{Q_L}, q^B_{L}, \Delta q^B)$,
for the particular choice $q_{Q_L}^B=-1, q_{L}^B=-1$, reproduces the entire charge assignment of a special class of intersecting brane models (see \cite{Ibanez:2001nd} and \cite{Ghilencea:2002da} and the discussion in \cite{Coriano:2009zh})
\bea
f(-1, -1, 4)=(-1, 0, 0, -1, 0, +2, -2).
\eea

In Fig. \ref{fig:gamma} (right panel) we show the decay rates as a function of the 
axion mass in both cases, having chosen a nominal mass range for this particle varying between $10^{-5} -1$
eV. One can immediately observe that the rates for the PQ case are smaller than those for the St\"uckelberg by a factor 
of $10^{20}-10^{12}$, nevertheless the axi-Higgs $\chi$ has a lifetime which is much bigger than the current age of the universe.

Concerning the possibility to detect the axion through its two-photon decay channel, its tiny mass and the smaller value of its lifetime unfortunately do not allow to set significant constraints on its possible parameter space. The situation, in this case, 
if rather different from that of other dark matter candidates, such as, for instance, the gravitinos, which have been widely investigated recently \cite{Bertone:2007aw,Choi:2009ng, Buchmuller:2007ui}. In fact, the allowed parameter space where the constraints derived from those previous studies apply, concern a region in the plane  ($\tau_{DM}, m_{DM}$) - with 
$\tau_{DM}$ being the lifetime of a generic dark matter particle and $m_{DM}$ its mass - which is bounded by the intervals 
$10^{26}\, \textrm{s} < \tau_{DM} < 10^{35}\, \textrm{s}$ and $ 10^{-5}\, \textrm{GeV} < m_{DM} < 10^{2} \,\textrm{GeV}$. 
While the value of $\tau_\chi$ for the axion can reasonably reach the lower edge of the scanned region in $\tau_{DM}$, by an adjustment of its coupling $g_B$ and charge assignments of the anomalous $U(1)$, its mass is definitely too small to be excluded by these types of analysis. These studies are, obviously, very interesting for candidates of heavier mass, such as gravitinos. Similar considerations apply in the case of LHC studies, given the small production rates for a very light axion. For much heavier axions, instead, these types of studies have been performed quite recently \cite{Coriano:2009zh}, but the behaviour of this particle, in this case, is akin a light Higgs rather than a long-lived light pseudoscalar. 

\section{Relic density at the electroweak and at the QCD phase transitions} 
In this section we proceed with the derivation of the relic densities for $\chi$ both at the electroweak and at the QCD phase transitions.

At the electroweak scale, we will assume that the flat direction parameterized by the St\"uckelberg axion $b$ is lifted by electroweak instanton corrections. A similar phenomenon, but much more sizeable, clearly will take place at the QCD phase transition.
As we have discussed previously, at the electroweak scale, a mixing between the various phases of the non-perturbative potential allows to identify the linear combination $\chi$ as the physical axion. In general, this is misaligned with respect to the minimum of the potential generated at this transition, with a misalignment that, as we have pointed out, is parameterized by the value $\theta= \chi/\sigma_\chi$. 

\begin{figure}[t]
\centering
\includegraphics[scale=.7]{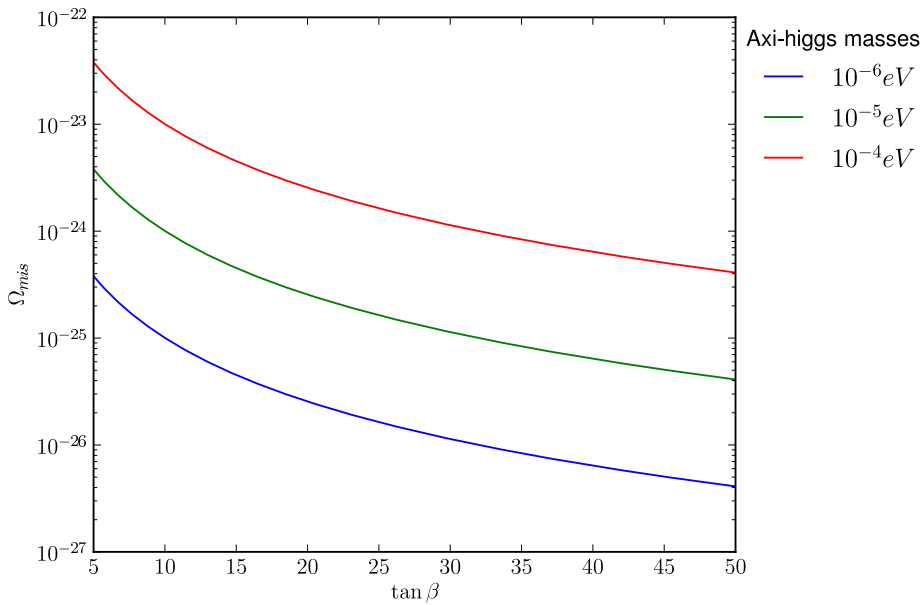}
\caption{Relic density of the axi-Higgs as a function of $\tan\beta$}
\label{fig:relicvu}
\end{figure}

The analysis of the relic density around the electroweak scale is then performed 
rather straightforwardly, following a standard approach borrowed from the PQ case. 
For this goal, we define the abundance variable of $\chi$

\begin{equation}
Y_\chi(T_i)\equiv \frac{n_\chi}{s}\bigg\vert_{T_i}
\label{abund}
\end{equation}
where $T_i$ is the oscillation temperature, which is close to the electroweak scale. 
The universe must be (at least) as old as the required period of oscillation in order for the axion field to start oscillating 
and to appear as dark matter, otherwise $\theta$ is misaligned but frozen. This is the content of the condition 
\begin{equation}
m_\chi(T_i)= 3 H(T_i),
\label{mhcond}
\end{equation}
between the mass of axion at the oscillation temperature $T_i$ ($m_\chi(T_i)$), and the Hubble parameter at the same temperature 
$H(T_i)$. 
The condition for oscillation Eq.~(\ref{mhcond}) allows to express the axion mass at $T=T_i$ in terms of the effective massless degrees of freedom evaluated at the same temperature, that is 
\begin{equation}
m_\chi(T_i)=\sqrt{\frac{4}{5}\pi^3 g_{*,T_i}}\frac{T_i^2}{M_P}.
\label{Tmass}
\end{equation}
Expressed in terms of the initial angle of misalignment $\theta_i$,  Eq.~\ref{abund} becomes  
\begin{equation}
Y_\chi(T_i)= \frac{45 \sigma_\chi^2\theta_i^2}{2\sqrt{5 \pi g_{*, T_i}} T_i M_P},
\label{ychi}
\end{equation}
where  $g_{*,T}=110.75$ is the number of massless degrees of freedom of the model at the electroweak scale.
Using the conservation of the abundance $Y_{a 0}=Y_{a}(T_i)$, the expression of the contribution to the relic density  is given by 
\begin{equation}
\Omega_\chi^{mis}=\frac{n_\chi}{s}\bigg\vert_{T_i} m_\chi\frac{s_0}{\rho_c}.
\label{omegaeq}
\end{equation}
The values of the critical energy density ($\rho_c$) and the entropy density today are estimated as
\begin{equation}
\rho_{c}=5.2\cdot10^{-6}\textrm{GeV}/\textrm{cm}^3\hspace{1cm}s_0=2970 \,\,\textrm{cm}^{-3},
\end{equation}
with $\theta\simeq1$. 
Given these values, the relic density as a function of $\tan\beta$ is given in Fig. \ref{fig:relicvu}. We have varied the oscillation mass and plotted the relic densities as a function of $\tan\beta$.  The variation of $v_u$ has been constrained  to give the values of the masses of the electroweak gauge bosons, via an appropriate choice of $\tan\beta$. 

For instance, if we assume a temperature of oscillation 
of $T_i=100$ \textrm{GeV}, an upper bound for the axi-Higgs mass, which allows the oscillations to take place, is $m_\chi(T_i)\approx 10^{-5} \textrm{eV}$, with $g_{*,T} \approx 100$. 

In order to specify $\sigma_{\chi}$ we have assumed a value of 1 TeV for the St\"uckelberg mass $M$, with $g_B\approx1$, and we have taken $(q_u,q_d)$ of order unity, obtaining $\sigma_{\chi}\simeq \,10^2$\,GeV.
As we lower the oscillation temperature (and hence the mass), the corresponding curves for $\Omega_\chi$ are down-shifted. 

The values of these relic densities at current time are basically vanishing and these small results are to be attributed to the value of $\sigma_\chi$, which is bound to vary around the electroweak scale. 

Just to compare with the PQ case, there $\sigma_\chi$ is replaced by the large scale $f_a$ at the QCD phase transition, and this is the reason of such a strong suppression for $\Omega_\chi$ (or of an enhancement, in the PQ case). Instanton effects at the electroweak scale are expected, in our case, to provide a mass of the type $m_\chi^2\sim\Lambda_{ew}^4/v^2$, with $\Lambda_{ew}^4\sim {\textrm {Exp}}(-2\pi/\alpha_w(v)) v^4$ - $\alpha_W(v)$ being the weak charge at the scale $v$ - which is indeed a rather small value since ${\textrm{ Exp}}(-2\pi/\alpha_w(v))\sim e^{-198}$. For this reason $\chi$ remains essentially a physical but frozen degree of freedom which may undergo a significant (second) misalignment only at the QCD phase transition. If not for the presence of a coupling of the axion to the gluons, via the color/ $U(1)_B$ mixed anomaly, $\chi$ could be classified as a quintessence axion, contributing to the dark energy content.

\subsection{The QCD phase transition}

We have seen that the electroweak phase transition has redefined the phase of $\chi$ via the mixing with the CP-odd Higgs sector, but below the electroweak scale the field remains essentially a pseudo Nambu-Goldstone mode which undergoes the second misalignment induced by the QCD phase transition, quite similarly to an ordinary PQ axion. Neglecting the small mass of $m_\chi$ induced at the electroweak scale, the new mass induced at 
the QCD scale is controlled by the ratio $m_\chi\sim\Lambda_{QCD}^2 v/M^2$, where now the angle of misalignment is essentially related to the St\"uckelberg mass $M$, via $M^2/v$ (which replaces $\sigma_\chi$), and is now given by $\theta^\prime\equiv\chi v/M^2$. 

In order to further clarify this point it is convenient to follow the analogy with the PQ case and observe that $M^2/v$ replaces $f_a$ in characterizing the coupling of the physical axion $\chi$ to the gluons. At the same time $M$, just like $f_a$, can be interpreted as a symmetry breaking scale, given the presence of a derivative coupling ($M\partial b\cdot B$) of the anomalous gauge boson $B$ to the St\"uckelberg field $b$ in the St\"uckeberg mass term. Thus it can be naturally interpreted, in this phase, as originating from a vev of an extra scalar singlet to which $B$ couples in the UV. These two elements clearly indicate that the new misalignment is basically given by $\theta'$. 

Given the similarity between this situation and the PQ case, then we can follow standard arguments to estimate the mass of $\chi$ after the QCD transition. Thus, we just recall that for PQ  axions \cite{Weinberg:1977ma,Visinelli:2009zm} the zero temperature mass is given by
\begin{equation}
m_a=N \frac{\sqrt{z}}{1+z}\frac{f_\pi m_\pi}{f_a}=6.2 \frac{N}{f_{a,12}} \mu\textrm{eV}
\end{equation}
where $z\simeq0.56$ is the ratio of the up and down quark masses, $f_\pi$ and $m_\pi$ are the pion decay constant and mass, $f_a$ is the axion decay constant, $f_{a,12}$ is the same constant expressed in units of $10^{12}\,\textrm{GeV}$ and N is the $U(1)_{PQ}$ color anomaly index. The dependence of the Peccei-Quinn axion on the temperature can be expressed as
\begin{equation}
m_a(T)=\left\{
\begin{tabular}{ll}
$m_a\,d\,\left(\frac{\Lambda}{T}\right)^4$ & $T>\Lambda$\\
$m_a$ & $T<\Lambda$,
\end{tabular}\right.
\label{moft}
\end{equation}
where $d$ is a model dependent numerical factor and $\Lambda_{QCD}\approx0.2\,\textrm{GeV}$ is the scale of the QCD phase transition. We have also set $d=0.018$ \cite{Turner:1985si}. 

We can borrow this formula to determine the mass of $\chi$ at zero temperature 
($m_\chi(T=0)\sim m_a$) and extend it to finite $T$  $(m_\chi(T)\sim m_a(T))$, using the same expression (\ref{moft}) valid in the case of the PQ axion. Coming to the value of the abundances, with the replacement of $\sigma\to M^2/v$, Eq.~\ref{ychi} assumes the form
\begin{equation}
Y_\chi(T_i)= \frac{45 \left(M^2/v\right)^2\left(\theta^\prime(T_i)\right)^2}{2\sqrt{5 \pi g_{*, T}} T_i M_P},
\end{equation}
being the exact analogue of the PQ expression for the abundances of the invisible axion, with $f_a \to M^2/v$. 
Concerning $g_{*,T}$, the effective massless degrees of freedom at $T\simeq1\,\textrm{GeV}$ are those of the gluons, the photon, 2 charged leptons, 3 neutrinos and 3 quark flavors; thus we have $g_{*,1GeV}=61.75$. At the QCD phase transition, that is at $T\simeq 200\,\textrm{MeV}$, the effective massless degrees of freedom are given by the photon, 1 charged leptons and 3 neutrinos, giving $g_{*,\Lambda}=10.75$.
The oscillation temperature can be obtained from Eq.~\ref{Tmass}
\begin{equation}
T_i^6=\sqrt{\frac{5}{4\pi^3\,g_{*,T_i}}}m_\chi(0) M_P\,b \Lambda_{QCD}^4,
\end{equation}
from which we get $T_i\approx0.6\,\textrm{GeV}$ and a typical oscillation mass which is given by $m_\chi(T_i)\approx1.4 \,\textrm{neV}$. 
Values of $m_\chi$ larger than this typical value will allow oscillations of the field $\chi$ and the appearance of relic densities whose size is essentially controlled by the value of $M$, the St\"uckelberg mass, via the scale $M^2/v$.

Given the analogy between $M^2/v$ and $f_a$ and the dependence of the axion field amplitudes on these two scales, it is natural to expect that only for large values of $M$ one should expect a significant contribution to the relic density of these new axions.

\begin{figure}[t]
\centering
\includegraphics[scale=.7]{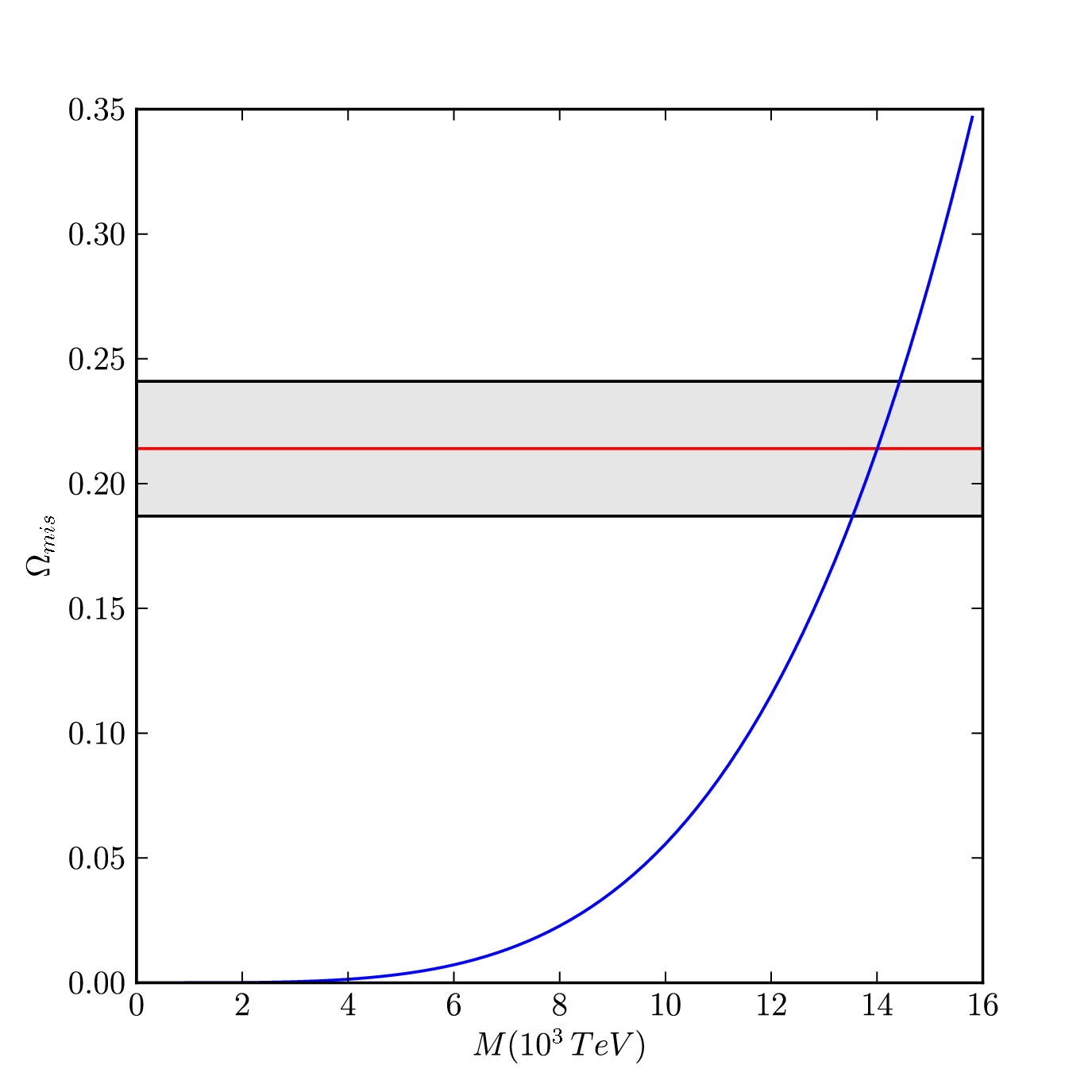}
\caption{Relic density of the axi-Higgs as a function of $M$. The grey bar represents the measured value of $\Omega_{DM} h^2=0.1123\pm0.0035$\label{fig:relicM}}
\end{figure}

We show in Fig. \ref{fig:relicM} results of a numerical study of $\Omega_{mis}h^2$ as a function of $M$, expressed in units of $10^9$ GeV. We show as a darkened area the bound coming from WMAP data~\cite{Jarosik:2010iu}, given as the average value plus an error band,  while the monotonic curve denotes the values of $\Omega_{mis}h^2$ as a function of $M$. 

It is clear that the relic density of $\chi$ can contribute significantly to the dark matter content only if the St\"uckelberg scale is rather large ($\sim 10^7$ GeV) and negligible otherwise. A final comment concerns the role of the isocurvature perturbations, which are generated by inflation, in these types of models, since in the case of the PQ axion they provide significant constraints on the possible values of $f_a$. The fact that the $b$ field does not correspond to a physical degree of freedom during inflation allows to bypass completely these constraints. They do not apply to these types of axions and this represents a very interesting feature and a significant variant of these models respect to the PQ case.

\section{Conclusions} 
We have discussed the most salient cosmological features of models containing gauged axions, obtained from the gauging of an anomalous symmetry.  The gauging allows to define a consistent theory for axion-like particles, which generalize many of the properties of PQ axions. They have appeared for the first time in the study of intersecting branes, but their features are quite generic. They are constructed as effective theories containing minimal gauge interactions which restore gauge invariance of the effective action in the presence of an anomalous $U(1)$ symmetry, and no further requirements. Differently from the PQ case, here there is no concept of an original PQ symmetry, broken at a very large scale, with the axion taking the role of a Goldstone mode that acquires a mass at the QCD phase transition. Rather, the physical axion emerges directly at the electroweak phase transition, when Higgs-axion mixing occurs. Being charged under $SU(3)$ and $SU(2)$, we have 
a sequential misalignment of this field, and we have quantified its relic density as a function of the St\"uckelberg mass. We have shown that only very large values of the St\"uckelberg mass cause a significant contribution of this type of axions to the current dark matter content of the Universe, which otherwise remains negligible. The absence of an original PQ-like potential has some implications at cosmological level, such as the absence of isocurvature perturbations, since the St\"uckelberg is not a physical mode before the electroweak phase transition, in particular at the time of inflation. This feature is due to the presence of a local gauge symmetry, realized in the St\"uckelberg form, which allows to absorb $b$  into the longitudinal component of the anomalous gauge boson. 

Our analysis represents, more generally, a description of the fate of the St\"uckelberg field in cosmology, from the defining St\"uckelberg phase of the theory at  a large scale (defined by the value of the St\"uckelberg mass), down to the electroweak and QCD phase transitions, when this field develops a physical component.  Our analysis could be extended in several directions, for instance with the inclusion of the modifications induced on the computation of the relics due to the presence of non-homogeneities in $\chi$ beyond the QCD horizon, a feature which is also present in PQ models when the PQ scale lays below the scale of inflation. However, even at this level of refinement, the only significant scale in the determination of the relic densities remains the value of the St\"uckelberg mass.  Small values of this mass parameter in the TeV range leave the contribution of these particles to the relic densities of dark matter negligible, and sizeable for $M$ around an intermediate scale of 
$10^7$ GeV. In this case all the constraints coming from the neutral current sector are satisfied, being the extra 
$Z^\prime$ of the theory completely decoupled from the low energy spectrum of the Standard Model. The appearance of this intermediate scale is a novel feature of this type of axions which could be used to 
set limits on their parameter space. 

\vspace{1cm}

\centerline{\bf Acknowledgements}
We thank Pierre Sikivie and Nikos Irges for discussions. C.C. thanks the Physics Department at Thessaloniki for hospitality. This work is supported in part  by the European Union through the Marie Curie Research and Training Network UniverseNet (MRTN-CT-2006-035863). 
\begin{appendix}

\section{Appendix. The model, definitions and conventions}
We summarize in this section some results concerning the model with a 
single anomalous $U(1)$ discussed in the main sections.

The effective action has the structure given by
\beqn
{\mathcal S} &=&   {\mathcal S}_0 + {\mathcal S}_{Yuk} +{\mathcal S}_{an} + {\mathcal S}_{WZ} + {\mathcal S}_{CS}
\label{defining}
\eeqn
where ${\mathcal S}_0$ is the classical action. It contains the usual gauge degrees of freedom of the Standard Model plus the extra anomalous gauge boson $B$ which is already massive, before electroweak symmetry breaking, via a St\"uckelberg mass term, reviewed in Sec.~\ref{sec:gener-feat-models}. Its complete expression is given in \cite{Coriano':2005js}. Here we briefly describe the structure of the anomalous contributions and of the induced counterterms for the restoration of gauge invariance in the 1-loop effective action.

In Eq.~(\ref{defining}) the anomalous contributions coming from the 1-loop triangle diagrams involving abelian and non-abelian gauge interactions are summarized by the expression
\beqn
{\mathcal S}_{an}&=& \frac{1}{2!} \langle T_{BWW} BWW \rangle +  \frac{1}{2!} \langle T_{BGG} BGG \rangle
 + \frac{1}{3!} \langle T_{BBB} BBB \rangle     \nonumber\\
&&+ \frac{1}{2!} \langle T_{BYY} BYY \rangle + \frac{1}{2!} \langle T_{YBB} YBB \rangle,
\eeqn
where the symbols $\langle \rangle$ denote integration. For instance, 
the contributions in configuration space are given explicitly by 
\ba
\langle T_{B W W} B W W \rangle&\equiv& \int dx\, dy \, dz
 T^{\lambda \mu \nu, ij}_{BWW}(z,x,y) B^{\lambda}(z) W^{\mu}_{i}(x)
W^{\nu}_{j}(y)
\ea
and so on, where $T_{BWW}$ denotes the anomalous triangle diagram with 
one $B$ field and two $W$'s external gauge lines. The gluons  are denoted by $G$.
The Wess-Zumino (WZ) counterterms are given by
\beqn
{\mathcal S}_{WZ}&=& {C_{BB}}\langle b\,  F_{B} \wedge F_{B}  \rangle
+ {C_{YY}}\langle b\, F_{Y} \wedge F_{Y}  \rangle + {C_{YB}} \langle b\, F_{Y} \wedge F_{B}  \rangle \nonumber\\
&&+ {F} \langle b\, Tr[F^W \wedge F^W]  \rangle  +  {D}   \langle b\, Tr[F^G \wedge F^G] \rangle,
\eeqn
while the gauge dependent Chern-Simons (CS) abelian and non abelian counterterms \cite{Anastasopoulos:2006cz} needed to cancel
the mixed anomalies involving a B line with any other gauge interaction of the
SM take the form
\beqn
{\mathcal S}_{CS}&=&+ d_{1} \langle BY \wedge F_{Y} \rangle + d_{2} \langle YB \wedge F_{B} \rangle  \nonumber\\
&&+ c_{1} \langle \epsilon^{\mu\nu\rho\sigma} B_{\mu} C^{SU(2)}_{\nu\rho\sigma} \rangle
+ c_{2} \langle \epsilon^{\mu\nu\rho\sigma} B_{\mu} C^{SU(3)}_{\nu\rho\sigma} \rangle.
\eeqn
The non-abelian CS forms given by
\beqn
C^{SU(2)}_{\mu \nu \rho} &=&  \frac{1}{6} \left[ W^{i}_{\mu} \left( F^W_{i,\,\nu \rho} + \frac{1  }{3} \, g^{}_{2}  
\, \varepsilon^{ijk} W^{j}_{\nu} W^{k}_{\rho}  \right) + cyclic   \right]              ,    \\
C^{SU(3)}_{\mu \nu \rho} &=&  \frac{1}{6} \left[ G^{a}_{\mu} \left( F^G_{a,\,\nu \rho} + \frac{1 }{3} \, g^{}_{3}
\, f^{abc} G^{b}_{\nu} G^{c}_{\rho}  \right) + cyclic  \right].
\eeqn

\begin{itemize}
\item{\bf The structure of $g^{\chi}_{\g\g}$}
\end{itemize}
The coefficients in front of the WZ counterterms are determined by requiring gauge invariance of the effective action.
We outline the case of  $g^{\chi}_{\g\g}$ and its relation to the fundamental parameters/scales of the theory. Among these are the St\"uckelberg mass $M$, the hypercharge and weak couplings $g_Y$ and $g_2$ and the charges of the fermion running inside the anomaly loops. These fix the coefficient of the anomalies $C_{BYY}$ and $C_{BWW}$ (for the $U(1)_B\, U(1)_Y ^2$ and $SU(2)^2\, U(1)_B$ anomalies) and the rotation matrices of the 
neutral gauge bosons $O^A$ and of the CP-odd sector $O^\chi$, defined in Eq.~(\ref{stella}). This is defined as in Eq.~(\ref{gchichi}) in terms of the counterterms
\begin{align}
F=  \frac{g^{}_{B}}{M}i  g^{2}_{2} \frac{a^{}_{n}}{2} C_{BWW},
\end{align}
with
\begin{equation}
C_{BWW} = - \frac{1}{8} \sum_{f} q_{fL}^{B},
\end{equation}
with $a_n=- \frac{i}{2 \pi^2}$ being the $AVV$ anomaly,
and
\begin{equation}
C_{YY} =  \frac{g_{B}}{M}  i  g^{\,2}_{Y}  \frac{a^{}_{n}}{2} C_{BYY},
\end{equation}
which is defined by the charges 
\begin{equation}
C_{BYY} = \frac{1}{8} \sum_{f} \left[ q_{fR}^{B} (q_{fR}^{Y})^{2}  - q_{fL}^{B} (q_{f_L}^{Y})^{2}  \right].
\end{equation}
Explicit expressions for $C_{BYY}$ and $C_{BWW}$ are given in Eq.~(\ref{charge_asynew}).

\begin{itemize}
\item{\bf Fermion interactions}
\end{itemize}
The covariant derivatives are defined as
\ba
D_\mu=\partial_\mu+i g_s T^a G^a_\mu+i g_2 \tau^a W^a_\mu+\frac{i}{2}g_Y q^Y Y_\mu+\frac{i}{2}g_B q^B B_\mu,
\ea
with $T^a$ and $\tau^a$ given by
\ba
T^a=\frac{\lambda^a}{2}\hspace{1cm}\tau^a=\frac{\sigma^a}{2},
\ea
where $\lambda^a$ and $\sigma^a$ are the Gell-Mann and Pauli matrices. This choice of the covariant derivative defines the gauge variations of the fields; in particular, under the abelian group transformations we have
\begin{equation}
B_\mu^\prime=B_\mu+\partial_\mu\theta \hspace{1cm} b^\prime=b+	M\theta \hspace{1cm}
\phi^\prime=e^{-i \frac{1}{2} g_B q^B_\phi \theta} \phi.
\end{equation}
We write the lepton doublet as
\ba
L_{i} = \begin{pmatrix} 
\nu_{L\,i} \cr 
e_{L\,i}
\end{pmatrix}
\ea
The interaction Lagrangian for the leptons becomes 
\begin{align}
{\cal L}_{int}^{lep} =& \begin{pmatrix}\bar{{\nu}}_{L\,i} & \bar{e}_{L\,i}\end{pmatrix}
\gamma^{\mu}
\left[-g_2 \tau^a W^a_\mu+\frac{1}{4}g_Y  Y_\mu-\frac{1}{2}g_B q^B_L B_\mu \right]  
\begin{pmatrix}
\nu_{Li} \cr 
e_{Li}
\end{pmatrix} +\nonumber\\
&\bar{e}_{R\,i}
\gamma^{\mu} 
\left[\frac{1}{2}g_Y Y_\mu-\frac{1}{2}g_B q^B_{e_R} B_\mu \right]e_{R\,i}.
\end{align}
As usual we define the left-handed and right-handed currents 
\beqn
J^{L}_{\mu} = \frac{1}{2}(J_{\mu} - J^{5}_{\mu}),  \qquad  J^{R}_{\mu} = \frac{1}{2}(J_{\mu} + J^{5}_{\mu}), 
\qquad J_{\mu} = J^{R}_{\mu} + J^{L}_{\mu}, \qquad  J^{5}_{\mu} = J^{R}_{\mu} - J^{L}_{\mu}.
\eeqn
Writing the quark doublet as
\bea
Q_{i} = \begin{pmatrix} 
u_{L\,i}\cr 
d_{L\,i} 
\end{pmatrix},
\eea
we obtain the interaction Lagrangian
\begin{align}
{\cal L}_{int}^{quarks} =&   
\begin{pmatrix} 
\bar{ u }_{L\,i} & \bar{ d }_{L\,i} 
\end{pmatrix}
\gamma^{\mu}
\left[-g_s T^a G^a_\mu- g_2 \tau^a W^a_\mu-\frac{1}{12}g_Y  Y_\mu-\frac{1}{2}g_B q^B_Q B_\mu \right]  
\begin{pmatrix} 
u_{L\,i} \cr 
d_{L\,i}
\end{pmatrix}  + \nonumber\\
& + \bar{u}_{R\,i} \gamma^{\mu} 
\left[-g_s T^a G^a_\mu- g_2 \tau^a W^a_\mu-\frac{1}{3}g_Y Y_\mu-\frac{1}{2}g_B q^B_{u_R} B_\mu \right]  u_{R\,i}  \nonumber\\
& + \bar{d}_{R\,i} \; {\gamma}^{\mu}
\left[-g_s T^a G^a_\mu- g_2 \tau^a W^a_\mu+\frac{1}{6}g_Y Y_\mu-\frac{1}{2}g_B q^B_{d_R} B_\mu\right] d_{R\,i}.
\end{align}
We work with a 2-Higgs doublet model, and therefore we parametrize the Higgs fields in terms of 8 real degrees of freedom as
\ba
H_u=\left(
\begin{array}{c}
H_u^+\\
H_u^0
\end{array}\right) \qquad 
H_d = \left(
\begin{array}{c}
H_d^+\\
H_d^0 
\end{array}\right)
\ea
where $H_u^+$, $H_d^+$ and $H_u^0$, $H_d^0$ are complex fields. Specifically
\ba
H_u^+ =  \frac{\textrm{Re}H_{u}^+ + i \textrm{Im}H_{u}^+}{\sqrt{2}} ,\qquad
H_d^- =  \frac{\textrm{Re}H_{d}^- + i \textrm{Im}H_{d}^-}{\sqrt{2}} , \qquad
H_u^- = H_u^{+ *}, \qquad
H_d^+ = H_d^{- *}.
\ea
Expanding around the vacuum we get for the uncharged components
\ba
H_u^0 =  v_u + \frac{\textrm{Re}H_{u}^0 + i \textrm{Im}H_{u}^0}{\sqrt{2}} , \qquad
H_d^0 =  v_d + \frac{\textrm{Re}H_{d}^0 + i \textrm{Im}H_{d}^0}{\sqrt{2}}. 
\label{Higgsneut}
\ea
The Weinberg angle is defined via $\cos\theta_W= g_2/g, \sin\theta_W= g_Y/g$, with $g^2= g_Y^2 + g_2^2$.
We also define $\cos \beta=v_d/v$, $ \sin \beta=v_u/v$ with $v^2=v_d^2 + v_u^2$.

\subsection{The Yukawa couplings and the axi-Higgs }
The couplings of the two Higgs and of the axi-Higgs to the fermion sector 
are entirely described by the Yukawa Lagrangian.
The Yukawa couplings of the model are given by
\bea
{\cal L}_{\rm Yuk}^{unit.} 
&=& - \Gamma^{d} \, \bar{Q} H_{d} d_{R} - \Gamma^{d} \, \bar{d}_R H^{\dagger}_{d} Q - 
\Gamma^{u} \, \bar{Q}_{L} (i \sigma_2 H^{*}_{u}) u_{R} 
- \Gamma^{u} \, \bar{u}_R (i \sigma_2 H^{*}_{u})^{\dagger} Q_{L} \nonumber\\
&&-   \Gamma^{e} \, \bar{L} H_{d} {e}_{R} - \Gamma^{e} \, \bar{e}_R H^{\dagger}_{d} L  \nonumber\\
&=& - \Gamma^{d} \, \bar{d} H^{0}_{d} P_{R} d - \Gamma^{d} \, \bar{d}  H^{0*}_{d} P_{L} d  
- \Gamma^{u} \, \bar{u}  H^{0*}_{u} P_{R} u - \Gamma^{u} \, \bar{u}  H^{0}_{u} P_{L} u \nonumber\\
&&- \Gamma^{e}  \, \bar{e} H^{0}_{d} P_{R} e - \Gamma^{e} \, \bar{e}  H^{0*}_{d} P_{L} e 	,
\label{yukawa_utile}
\eea
where the Yukawa coupling constants $\Gamma^{d}, \Gamma^{u}$ and $ \Gamma^{e}$  run over the three generations, i.e. $u = \{u, c, t\}$, $d = \{d, s, b\}$ and $e$ = \{$e$, $\mu$, $\tau$\}.
Rotating the CP-odd and CP-even neutral sectors into the mass eigenstates and 
expanding around the vacuum  we obtain 
\begin{align}
H_u^0 =& v_u + \frac{  Re{H^0_{u}} + i \, Im{H^0_u}}{\sqrt{2}}  \nonumber\\
=&  v_u + \frac{  (h^0 \sin\a  - H^0 \cos\a ) 
+ i \, \left(O^{\chi}_{11}G^1_0 + O^{\chi}_{21}G^2_0 + O^\chi_{31} \chi	  \right) }{\sqrt{2}}  
\label{Higgs_up}   
\end{align}
\begin{align}     
H_d^0 =&  v_d + \frac{ Re{H^0_d} + i \, Im{H^0_d}}{\sqrt{2}}   \nonumber\\
=&  v_d  +  \frac{  (h^0 \cos\alpha  + H^0 \sin\alpha ) 
+ i \left( O^{\chi}_{12}G^1_0 + O^{\chi}_{22}G^1_0 + O^\chi_{32} \chi \right) }{\sqrt{2}}  
\label{Higgs_down}
\end{align}
so that in the unitary gauge we obtain
\begin{align}
H_u^0 =& v_u + \frac{1}{\sqrt{2}} \left[ (h^0 \sin{\a}  - H^0 \cos{\a})  +
i \, O^{\chi}_{31} \chi \right] \nonumber\\
H_d^0 =& v_d + \frac{1}{\sqrt{2}} \left[ (h^0 \cos{\a}  + H^0 \sin{\a})  +
i  O^{\chi}_{32}\chi \right],  
\label{Higgsdec}
\end{align}
where the vevs of the two neutral Higgs bosons $v_u=v \sin \beta $ and $v_d= v \cos \beta $ satisfy 
\ba
\tan \beta = \frac{v_u}{v_d}, \qquad  v=\sqrt{v_u^2+v_d^2}.
\ea
The fermion masses are given by
\ba
&& m_{u} =  {v_u \G^{u}},\hskip 1cm  m_{\n } =  {v_u \G^\n},  \nonumber\\
&& m_{d} =  {v_d \G^d},\hskip 1cm  m_{e} =  {v_d \G^e},
\label{f_masses}
\ea
where the generation index has been suppressed for brevity. The fermion masses, defined in terms of the two expectation values $v_u,v_d$ of the model, show an enhancement of the down-type Yukawa couplings for large values of $\tan \beta$ while at the same time the up-type Yukawa couplings get a suppression. The couplings of the $h^0$ boson to fermions are given by
\ba
{\cal L}_{\rm Yuk}(h^0) =  -  \Gamma^d \, \bar{d}_{L} d_R
\left( \frac{ \cos\a}{\sqrt{2}} h^0 \right) - \Gamma^u \, 
\bar{u}_{L} u_{R}   \left( \frac{ \sin\a }{\sqrt{2}} h^0  \right)  
-  \Gamma^e \, \bar{e}_{L}e_R  \left( \frac{ \cos\a}{\sqrt{2}} h^0 \right)  + c.c. 
\ea
The couplings of the $H^0$ boson to the fermions are 
\ba
{\cal L}_{\rm Yuk}(H^0) =  - \Gamma^d \, \bar{d}_{L} d_R 
\left( \frac{ \sin\a}{\sqrt{2}} H^0 \right) -   \Gamma^u \, 
\bar{u}_{L} u_{R}   \left( - \frac{ \cos\a }{\sqrt{2}} H^0  \right)
 - \Gamma^e \, \bar{e}_{L}e_R  \left( \frac{ \sin\a}{\sqrt{2}} H^0 \right) + c.c. 
\ea
The physical gauge fields can be obtained from the rotation matrix $O^A$
\ba
\begin{pmatrix}A_\g \cr Z \cr Z^{\prime} 
\end{pmatrix}\,=O^A \begin{pmatrix} W_3 \cr A^Y \cr B
\end{pmatrix}
\label{OA}
\ea
which can be approximated at the first order as
\bea
O^A  \simeq  \begin{pmatrix}
\frac{g^{}_Y}{g}           &     \frac{g^{}_2}{g}         &      0   \cr
\frac{g^{}_2}{g} + O(\epsilon_1^2)          &     -\frac{g^{}_Y}{g} + O(\epsilon_1^2) &      \frac{g}{2} \epsilon_1 \cr
-\frac{g^{}_2}{2}\epsilon_1     &     \frac{g^{}_Y}{2}\epsilon_1  &   1 + O(\epsilon_1^2)
\end{pmatrix}
\label{matrixO}
\eea
where
\begin{align}
&\epsilon_1=\frac{x_B}{M^2},\nonumber\\
&x_B=\left(q^{B}_u v_u^2 + q^{B}_d v_d^2\right).
\end{align}
More details can be found in \cite{Coriano:2007xg}.

\end{appendix}

\end{document}